\documentclass[preprint,12pt]{elsarticle}
\usepackage{natbib}

\usepackage{amssymb}
\usepackage{amsmath}

\usepackage{graphicx}
\usepackage[hidelinks]{hyperref}
\hypersetup{
  colorlinks=true,
  linkcolor=blue,
  citecolor=black,
  filecolor=magenta,
  urlcolor=blue, }

\usepackage{amsmath,amssymb,amsfonts}
\usepackage{xcolor}
\usepackage{tcolorbox}
\usepackage{float}
\usepackage{adjustbox}
\usepackage{etc}
\usepackage{soul}
\usepackage{caption}
\usepackage{subcaption}
\usepackage{mwe}
\usepackage{tcolorbox}
\usepackage{enumerate}
\usepackage{makecell}
\usepackage[super]{nth}
\usepackage{algorithm}
\usepackage{algpseudocode} 
\usepackage{enumitem}

\usepackage{graphicx}
\usepackage{textcomp}
\usepackage{xcolor}
\usepackage{tcolorbox}
\usepackage{float}
\usepackage{adjustbox}
\usepackage{etc}
\usepackage{tabularx} 
\usepackage{array}
\usepackage{moreverb}
\usepackage{multirow}
\usepackage{caption}
\usepackage{tikz}
\usepackage{longtable}
\usepackage{arydshln}
\usepackage{dashrule}

\definecolor{lightergray}{RGB}{220, 220, 220}
\newcommand{\bluetext}[1]{\textcolor{blue}{#1}}

\journal{Journal of Systems and Software}
\date{}
\begin{document}

\begin{frontmatter}



\title{Improving IR-based Bug Localization with Semantics-Driven Query Reduction}



\author{Asif Mohammed Samir}
\author{Mohammad Masudur Rahman}

\affiliation{organization={Dalhousie University}, 
            city={Halifax}, 
            state={NS}, 
            country={Canada}}

\begin{abstract}
Despite decades of research, software bug localization remains challenging due to heterogeneous content and inherent ambiguities in bug reports. Existing methods, such as Information Retrieval (IR)-based approaches, often attempt to match source documents to bug reports, overlooking the context and semantics of the source code. On the other hand, Large Language Models (LLMs) (e.g., Transformer models) show promising results in understanding both texts and code. However, they have not yet been adapted well to localize software bugs using bug reports. They could also be data or resource-intensive. To bridge this gap, we propose, \textit{IQLoc}, a novel approach that capitalizes on the strengths of both IR and LLMs for bug localization. In particular, we leverage the transformer-based model's understanding of code semantics to reason about its suspiciousness and to reformulate search queries and thus enhance bug localization using Information Retrieval. To evaluate IQLoc, we refine the Bench4BL benchmark dataset and extend it by incorporating $\approx$30\% more recent bug reports, resulting in a benchmark containing $\approx$7.5K bug reports. We evaluated IQLoc using three performance metrics and compare it against eight baseline techniques. Experimental results demonstrate its superiority, achieving up to 100.40\% and 78.08\% in MAP, 61.49\% and 64.58\% in MRR, and 76.98\% and 100.90\% in HIT@K for the test bug reports with random and time-wise splits, respectively. Moreover, IQLoc improves MAP by 118.70\% for bug reports with stack traces, 111.87\% for those that include code elements, and 127.45\% for those containing only descriptions in natural language. By integrating program semantic understanding into Information Retrieval, IQLoc mitigates several longstanding challenges of traditional IR-based approaches in bug localization.
\end{abstract}



\begin{keyword}


Bug Localization \sep Hybrid \sep Information Retrieval \sep LLM \sep Program Semantics
\end{keyword}

\end{frontmatter}

\section{Introduction}
\label{sec:intro_iploc}

Software maintenance claimed \$2.4 trillion in 2022 from the US economy, a $\approx$84\% increase over the previous two years \cite{US_maintenance_stats_2022}. A significant factor contributing to this increase in costs is software bug, which prevents a software system from working correctly \cite{SE_Terminology}. These bugs not only hinder the functionality of software but also can lead to severe system failures (e.g., AT\&T Mobility outage)\cite{FCC2024_ATnT, Reuters2024_ATnT, Summary_ATnT, nasaMariner}. Consequently, software developers dedicate 35-50\% of their time tackling these bugs \cite{o2017debugging_stats, britton2013reversible_stats}. Zou et al. \cite{practitioners_bug_localization_study} recently conduct a practitioner survey where they select ten different tasks related to software bugs and collect responses from 327 software practitioners (e.g., developers, project managers, testers) affiliated with leading IT companies (e.g., Google, Meta, Microsoft, Amazon). According to their survey, 82.4\% of the practitioners consider bug localization as the most important or an important task. The task involves identifying the specific locations in the software code responsible for a software bug or failure.   

During bug localization, developers often rely on bug reports to find the locations of problematic code. However, understanding bugs from their description in the reports and locating them in the software code is challenging and time-consuming, even for seasoned developers \cite{rahman2021forgotten}. Bug reports contain natural language text and program artifacts such as stack traces, program elements, and code diffs \cite{blizzard}. The presence of these diverse contents in bug reports and the inherent ambiguities in natural language text \cite{furnas1987vocabulary} can significantly complicate the process of bug localization. To address these challenges, researchers have been actively pursuing automated solutions for bug localization over the last 50 years \cite{ir_bug_localization1, ir_bug_localization2_spectra, ir_bug_localization3_topic, ir_bug_localization4,dnnloc_ir, ir_localization_lda_buglocator, saha_bleuir, ir_ml_amalgam}.

Over the last few decades, many methods have been employed to automatically localize software bugs. One frequently used method is Information Retrieval (IR). Existing IR-based approaches \cite{ir_bug_localization1, ir_bug_localization2_spectra, ir_bug_localization3_topic, ir_bug_localization4} often use pre-processed texts from bug reports as search queries and attempt to match them with source code, overlooking the context or semantics of the code. This can lead to spurious matching between bug reports and source code, given the variability of natural language text describing the software bug \cite{query_keyword, furnas1987vocabulary}. Several existing approaches incorporate code change history, version control history, or even code authoring history to improve bug localization \cite{bl_code_change_BLIA, ir_ml_amalgam}. Although additional contexts of a software bug are captured, these approaches rely primarily on statistical chances rather than a comprehensive understanding of the program semantics, limiting their effectiveness in bug localization. A recent work \cite{bench4bl} shows that these techniques perform comparably despite incorporating additional contextual information. Furthermore, another recent work \cite{rahman2021forgotten} suggests that the performance of existing IR-based techniques might be significantly affected by their selected queries from the bug reports.

Recently, several approaches have focused on making appropriate search queries based on bug reports to improve bug localization. Rahman and Roy \cite{strict} leverage both co-occurrences and syntactic dependencies among words from a bug report to capture meaningful search queries. In a recent work \cite{blizzard}, the authors also show the use of static and hierarchical dependencies among program elements to construct queries. Other techniques \cite{chaparro2017_ob_eb, sisman2012incorporating} make queries from a bug report through keyword expansion or noise removal. Although effective in making queries, these techniques could be limited by the vocabularies of bug reports, which are of varying quality \cite{rahman2021forgotten}. This limitation restricts their capacity to comprehend contextual intricacies and locate defective source code documents. In contrast, deep learning-based techniques show promise in understanding the conceptual relationship between code and text \cite{cao2020bugpecker, zhu2020cooba, xiao2023bugradar}. Large Language Models (LLMs) advance this capability by leveraging transformer architectures and self-attention mechanisms to capture nuanced semantic relationships. Being trained on corpora beyond bug reports, these techniques can draw inferences from a larger knowledge base. However, they fail to leverage the benefits that the existing methods (e.g., IR-based) have to offer during bug localization. Moreover, they require extensive data and computational power, hindering their large-scale, sustainable adoption.

To design a comprehensive solution capitalizing on the strengths of both approaches above, 
we propose
\textit{IQLoc}, a novel hybrid approach combining Information Retrieval (IR) with transformer-based models to localize software bugs. Textual relevance alone may broaden a search space, potentially delivering unrelated results. Our approach narrows down the search space by leveraging program semantics understanding of the pre-trained language models during bug localization. First, IQLoc retrieves the top K source documents textually relevant to a bug report (a.k.a., query) using an IR-based approach (e.g., the BM25 algorithm \cite{robertson1995okapi_bm25}). Second, it trains a transformer-based, cross-encoder model to assess their relevance (based on their program semantics) to the bug report and narrows down the search space. Finally, IQLoc reduces the search query leveraging the above documents, reranks the documents using the query, and returns the suspicious source code documents.

We selected an existing benchmark dataset -- Bench4BL \cite{bench4bl} -- for our experiments and extended it with recent bug reports (i.e., submitted until September 2024). To refine the dataset, we selected the bug reports that include version information and have corresponding relevant documents in their respective GitHub repositories. We further expanded it with $\approx$30\% more recent bug reports, resulting in a final set of 7,483 bug reports. To evaluate our proposed technique, we employed three appropriate and widely used metrics: Mean Average Precision (MAP), Mean Reciprocal Rank (MRR), and HIT@K. We compare our technique with eight appropriate baselines including -- BLUiR \cite{saha_bleuir}, Blizzard \cite{blizzard}, BRTracer \cite{BRTracer}, BLIA \cite{bl_code_change_BLIA}, LRBL \cite{LRBL_learning2rank}, LLmiRQ \cite{LLmiRQ}, DNNLoc \cite{dnnloc_ir} and RLocator \cite{RLocator}. Across various measures, IQLoc consistently outperformed these techniques, with improvements of up to 100.40\% and 78.08\% in MAP, 61.49\% and 64.58\% in MRR, and 76.98\% and 100.90\% in HIT@K for the test bug reports with random and time-wise splits, respectively. Additionally, IQLoc improves bug localization over baselines in terms of MAP and MRR, achieving 118.70\% and 86.49\% for bug reports with stack traces, 111.87\% and 71.09\% for those containing program artifacts, and 127.45\% and 75.23\% for natural language-only bug reports.


Thus, this research makes the following contributions:

\begin{itemize}
    \item A novel hybrid bug localization technique, \textit{IQLoc}, that capitalizes on the strengths of both traditional (e.g., query reformulation, IR) and deep learning-based approaches (e.g., transformer models) and leverages textual relevance, program semantics relevance, and language models in bug localization.
    \item A refined and extended Bench4BL dataset incorporating recent bug reports, resulting in $\approx$7.5K bug reports.
    \item A comprehensive evaluation of IQLoc using three popular metrics, two different splits of the dataset, and comparison with eight baseline techniques \cite{saha_bleuir, blizzard, BRTracer, bl_code_change_BLIA, LRBL_learning2rank, LLmiRQ, dnnloc_ir, RLocator} from the literature.  
    \item A replication bundle\footnote{https://github.com/asifsamir/IQLoc} comprising a prototype, a carefully curated experimental dataset, configuration details, and trained and pre-trained models for third-party use and replication.
\end{itemize}

.
\begin{table}[!h]
\caption{A Motivational Example Illustrating Bug Localization Techniques and Approaches (Spring Webflow \# SWF-1416)}
\label{tab:bug_motivational_example}
\vspace{-0.5\baselineskip}
\centering

{\fontsize{7}{8}\selectfont
\resizebox{0.9\columnwidth}{!}{%
  \begin{tabular}{|p{8cm}|}
    \hline
    \rule{0pt}{2.5ex}\textbf{Title:} Turn off snapshot creation when max-snapshots = 0 \\[2pt]
    \textbf{Description:} Creating snapshots involves serializing and compressing the flow execution object, which can lead to issues with some PersistenceContext providers. Having the ability to turn off snapshot creation will provide an option when such issues occur at the cost of losing back button support.\rule[-1.5ex]{0pt}{0pt}\\
    \hline
  \end{tabular}
}}\\ \vspace{0.3em}

{\fontsize{7}{8}\selectfont
\resizebox{0.9\columnwidth}{!}{%
  \begin{tabular}{|p{2.2cm}|p{3.8cm}|p{0.8cm}|}
    \hline
    \textbf{Technique} & \textbf{Approach} & \textbf{Rank} \\ \hline

    Bug Title (Query)
    & Traditional IR (Bug Title Only)
    & 72 \\ \hline

    Bug Description (Query)
    & Traditional IR (Bug Description Only)
    & 57 \\ \hline

    Baseline Query
    & Traditional IR (Bug Title + Bug Description)
    & 53 \\ \hline

    Blizzard~\cite{blizzard}
    & Traditional IR (Context-Aware Query Reformulation)
    & 18 \\ \hline

    LRBL~\cite{LRBL_learning2rank}
    & Learning-to-Rank (Feature-Based Ranking)
    & 19 \\ \hline

    DNNLoc~\cite{dnnloc_ir}
    & Deep Learning (Neural Ranking Model)
    & 28 \\ \hline

    \bluetext{RLocator}~\cite{RLocator}
    & Reinforcement Learning (Ranking Optimization)
    & 6 \\ \hline

    IQLoc (Proposed)
    & Program Semantic Understanding
    & \textbf{1} \\ \hline
  \end{tabular}
 }}\\ 

\end{table}

\begin{figure}[!h]
    \centering
    \begin{tabular}{c} 
        \includegraphics[width=0.75\linewidth]{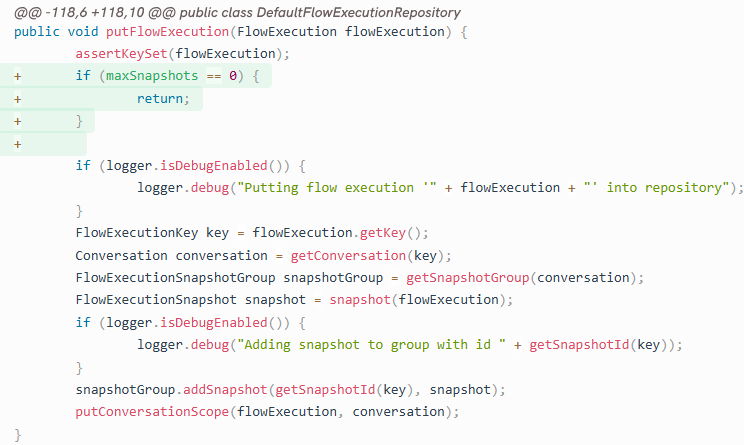} 
        
    \end{tabular}
    \caption{Buggy Code and Method Context}
    \label{fig:bugreport_method_context}
\end{figure}

\section{Motivational Example}
\label{sec:mot_example_iqloc}
In this section, we demonstrate the effectiveness of our approach --IQLoc-- in bug localization using a motivating example. We present an example where our approach outperforms the baseline methods. Table \ref{tab:bug_motivational_example} shows an example bug report (Bug ID\#SWF-1416) from \textit{Spring Webflow (SFW)}~\cite{springwebflow} and Fig. \ref{fig:bugreport_method_context} shows the associated buggy code. When the pre-processed version of the bug report is executed as a query by an IR-based search engine (e.g., Elasticsearch), the first buggy document is found at the \nth{53} position, which is not an ideal result. As shown in Table \ref{tab:bug_motivational_example}, the pre-processed version of the title or description field also does not make a good query.

 The example bug report discusses an issue involving snapshot creation, which could affect several classes including \texttt{PersistentContext}. Interestingly, the root cause of the bug involves the serialization/compression of the \texttt{FlowExecution} object. However, traditional techniques such as BLUiR and Blizzard consistently select \texttt{PersistenceContext} as a query keyword since it is found in the bug report. This leads to poor retrieval performance and they retrieve their first buggy documents at the \nth{13} and \nth{18} positions, respectively. 

In contrast, LRBL, a learning-to-rank technique, leverages a mix of heterogeneous features including lexical similarity, class-name overlap, collaborative filtering, and historical bug-fixing information to rank suspicious source documents. Similarly, DNNLoc, the seminal work on deep learning–based bug localization, also considers multiple features (e.g., textual relevance, change history, and collaborative filtering) and employs a neural ranking model to predict suspiciousness scores. Nevertheless, LRBL and DNNLoc perform poorly for this bug, retrieving the buggy document at the \nth{19} and \nth{28} positions, respectively. On the other hand, RLocator, a reinforcement learning–based technique, formulates bug localization as a ranking optimization problem. However, it still fails to capture the nuances of program understanding and ranks the buggy document at the \nth{6} position, which is not ideal.

In comparison, IQLoc attempts to understand the context of a reported bug from the source code. In particular, its cross-encoder module learns to connect bug reports to their corresponding buggy code through transformer-based, contextual learning. The cross-encoder module of our technique was able to rank the buggy code at the \nth{11} position. Our module achieves that by capturing the context of the bug from the bug report and making a connection with the intent of the buggy code (e.g., snapshot creation from \texttt{FlowExecution}). More interestingly, upon reformulating the query (a.k.a., preprocessed bug report) leveraging the IR-based and cross-encoder-based components, our technique delivers the buggy source document (e.g., Fig. \ref{fig:bugreport_method_context}) at the \nth{1} position, which is ideal and highly promising.

\section{Background}
\label{sec:background_iqloc}

\subsection{Program Semantics and Context}
 Program semantics refer to the meaning or behavior of a computer program, explaining how it processes inputs, performs computations, and produces outputs \cite{semantics_programming_lang}. They provide a structured way to reason about a program’s functionality, correctness, and execution behavior \cite{semcoder_program_semantics}. However, program semantics cannot be fully understood in isolation — they require context. Variables, functions, and control flows must be interpreted by capturing their surrounding program elements. Existing literature \cite{context_definition} suggests that context is not a static property but rather something that emerges dynamically from the interaction and relationships among program elements. To localize software bugs, we need to analyze not only the source code but also its semantics within its context. In particular, we leverage the Large Language Models' ability to understand program semantics in localizing software bugs.

\begin{figure}[t]
    \centering
    \begin{tabular}{c} 
        \centering
        \includegraphics[width=0.5\linewidth]{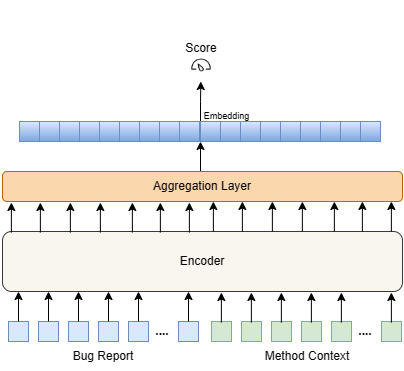}
    \end{tabular}

    \caption{Cross-Encoder's Architecture}
    \label{fig:cross-encoders}
\end{figure}

\subsection{{Cross-Encoders}}
\label{sec:background_cross_encoder}
The cross-encoder \cite{polyencoder} is a transformer-based approach used for reranking documents by enabling deep interactions between the input context and the candidate. This approach processes the context and the candidate jointly within a single transformer model. The inputs are concatenated with a special separator token (e.g., \texttt{[SEP]}) and passed through the transformer, allowing the self-attention mechanism to capture fine-grained dependencies between them (Fig. \ref{fig:cross-encoders}). 

Given an input context \texttt{ctxt} and a candidate \texttt{cand}, the transformer \(T\) produces a joint representation, typically extracted from the first token of the output sequence:

\[
y_{\text{ctxt,cand}} = h_1 = \text{first}(T(\text{ctxt}, \text{cand}))
\]

During encoding, the candidate label can focus on specific features within the input context, enabling the model to identify the most relevant input features for each candidate. This property makes the cross-encoder particularly effective for reranking tasks where fine-grained relevance is warranted.

Prior studies \cite{crossencoder_example_1, crossencoder_example_2} have demonstrated the cross-encoder’s advantages in refining retrieved results by analyzing relationships between queries and documents. Therefore, we use a cross-encoder in IQLoc to rerank source documents based on program semantic relevance between source code and bug reports.


\section{Methodology}
\label{sec:methodology_iqloc}

\begin{figure}[h]
    \centering
    \hdashrule{0.9\linewidth}{0.5pt}{2pt 2pt}  
    
    \vspace{2pt}
    \includegraphics[width=0.9\linewidth]{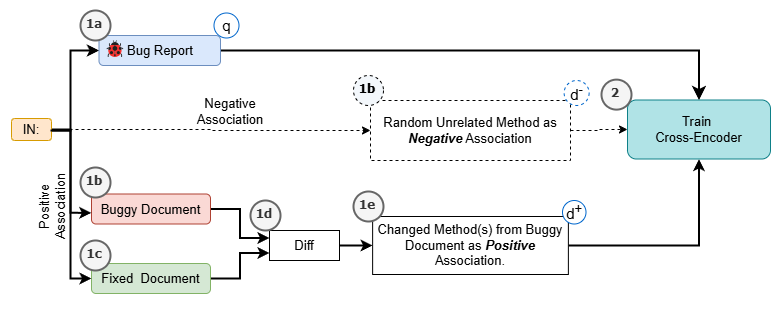}
    
    \vspace{2pt}
    {\small (a) Fine-tuning}
    
    \vspace{4pt}
    \hdashrule{0.9\linewidth}{0.5pt}{2pt 2pt}
    
    \vspace{2pt}
    \includegraphics[width=0.65\linewidth]{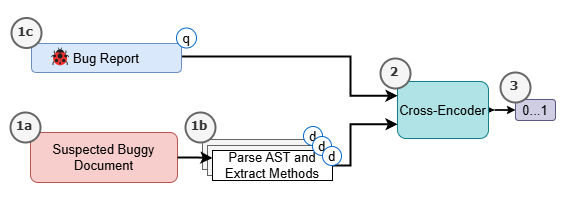}
    
    \vspace{2pt}
    {\small (b) Inference}
    
    \vspace{4pt}
    \hdashrule{0.9\linewidth}{0.5pt}{2pt 2pt}
    
    \caption{Fine-tuning and Prediction of Cross-Encoder Model}
    \label{fig:fine-tuning_prediction_CE}
\end{figure}

In this section, we present the methodology of our proposed technique --IQLoc-- for software bug localization. First, we describe the fine-tuning process of the Cross-Encoder model, shown in Fig. \ref{fig:fine-tuning_prediction_CE}, which assesses the relevance between bug reports and source documents. Then, we explain how the fine-tuned model is integrated into the overall bug localization workflow, as illustrated in Fig. \ref{fig:schematic-workflow-diagram}.

\subsection{Fine-tune Cross-Encoder Model}
\label{section:Finetune_CE}

\begin{figure}
    \centering
    \includegraphics[width=1.05\linewidth]{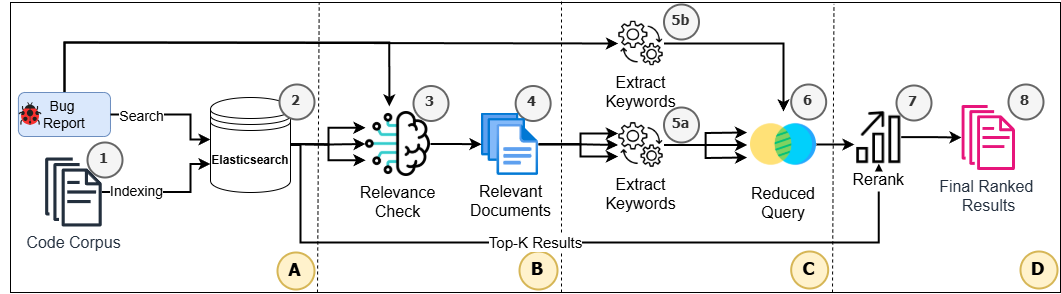}
    \caption{\centering Schematic Diagram of \textit{IQLoc}: \textit{(A)} Indexing \& Retrieval of Documents, \textit{(B)} Check Relevancy, \textit{(C)} Query Reformulation, \textit{(D)} Bug-Localization} 
    \label{fig:schematic-workflow-diagram}
\end{figure}

In IQLoc, we employ a cross-encoder model to determine relevance between a bug report and a code segment (Step 3, Fig. \ref{fig:schematic-workflow-diagram}). First, we fine-tune the base model of CodeBERT, a transformer-based encoder model \cite{feng2020codebert}, using bug-fix changes and enable it to differentiate between buggy and non-buggy instances during inference (Fig. \ref{fig:fine-tuning_prediction_CE}(a)). In particular, we append feed-forward network to the pre-trained CodeBERT model to achieve classification. Due to its pre-training on a large amount of natural language text and source code, the model is well-suited for our task involving bug reports and source code documents.

To fine-tune the model, we first parse each source code document from the training corpus and extract its buggy methods by leveraging the bug-fix diffs from corresponding version control history (Step 1c, Fig. \ref{fig:fine-tuning_prediction_CE}(a)). We detailed this method extraction process in Section \ref{section:dataset_constraction}. We also establish connection between bug reports and corresponding buggy code using appropriate heuristics \cite{bench4bl}. Then we treat bug reports (Step 1a) and buggy code pairs (Step 1e) as \textit{positive} samples in our training dataset. These pairs are fed into the cross-encoder model for training with the goal of establishing \textit{positive} contextual associations between buggy methods and their respective bug reports (Step 2, Fig. \ref{fig:fine-tuning_prediction_CE}(a)).

To train the model to differentiate between buggy and non-buggy samples, we also add negative pairs where each pair consists of a bug report (Step 1a, Fig. \ref{fig:fine-tuning_prediction_CE}(a)) and a randomly selected method body (Step 1b-dashed) to distinguish buggy from non-buggy code by leveraging program semantics. The capability stems from CodeBERT’s pre-training on a large bimodal corpus, which equips it with rich semantic representations of code structure and intent \cite{feng2020codebert}, as evidenced by its performance on downstream tasks (e.g., code summarization, clone detection \cite{codeClone1CodeBert, codeClone2ProbeCodeBert, programTranslationCodeBert}). Fine-tuning further adapts these representations by reinforcing links between real-world bug reports and their corresponding buggy code, enabling the model to capture subtle semantic cues associated with buggy behavior.

 To fine-tune our cross-encoder model, we configure the hyperparameters following standard practices \cite{feng2020codebert}. We set our initial learning rate to $1\times10^{-4}$ and optimized the model automatically using PyTorch’s \texttt{ReduceLROnPlateau} scheduler~\cite{pytorch}, which halved the rate whenever the validation loss did not improve for consecutive epochs (e.g., 2). The batch size was set to 8 to balance memory usage and gradient stability, given the computational constraints of our cloud server. We set the number of epochs at 100 at the beginning of the tuning; however, the effective training duration was determined dynamically, as the scheduler progressively reduced the learning rate and training stabilized once the loss plateaued. This configuration aligns with widely adopted fine-tuning strategies in the literature, which emphasize adaptive scheduling and validation-based stopping to ensure stable convergence and prevent overfitting~\cite{devlin2019bert_HypParam}.

\subsection{Corpus Indexing}
As we aim to localize software bugs using an Information Retrieval (IR)-based solution, it is essential to index all the source documents of a code base (a.k.a., corpus). In our approach, we chose Elasticsearch \cite{elasticsearch} due to its efficient handling of diverse data types, seamless integration with research tools, and robustness in document indexing. We index all source code (Step 1, Fig. \ref{fig:schematic-workflow-diagram}) from the 1,578 
 buggy versions of 42 subject-systems from the Bench4BL dataset.

\subsection{Retrieval of Potentially Buggy Source Documents}
\label{elasticsearch_result}
We use Elasticsearch to retrieve potentially buggy source documents by executing queries selected from bug reports (Step 2, Fig. \ref{fig:schematic-workflow-diagram}). Elasticsearch employs its default Standard Analyzer to preprocess queries and retrieve the top-K results (e.g., K = 100) from the corpus. During this retrieval process, we use the default scoring function of Elasticsearch, Okapi BM25 \cite{robertson1995okapi_bm25}, to assess the relevance of the search results. When retrieving suspicious buggy documents, we specify the particular project and version of the bug report where the bug occurred. Unlike other bug localization techniques \cite{saha_bleuir, ir_localization_lda_buglocator, blizzard}, which consider only the most recent version of a project and may retrieve unrelated documents, our approach ensures that we localize bugs within the correct historical context. This allows us to better mimic real-world scenarios by retrieving documents from the appropriate project and version. At the end of this step, we get a list of potential buggy source documents based on their textual relevance to the bug report.

\subsection{Relevance Estimation using Cross-Encoder}
After obtaining results from Elasticsearch, we employ our fine-tuned cross-encoder model to estimate their program semantic relevance to the bug report (Step 3, Fig. \ref{fig:schematic-workflow-diagram}).  
Given that we trained the cross-encoder model to differentiate between buggy and bug-free code (Fig. 3.2), its reasoning against the candidate methods should be useful. 
While source code vocabulary is important, our objective is to leverage the cross-encoder's joint encoding capability (Section \ref{sec:background_cross_encoder}) and capture deep semantic interactions between a bug report and a candidate code through an attention mechanism. Unlike independently learned representations (e.g., embedding-based similarity \cite{yang_word_embedding, polyencoder}), our approach enables us to capture program semantics based on fine-grained, token-level interactions.

 We analyze each retrieved source code document by parsing its Abstract Syntax Tree (AST) and extracting its individual methods (Step 1a-1b, \ref{fig:fine-tuning_prediction_CE}(b)). Each method is then paired with the bug report (Step 1c, \ref{fig:fine-tuning_prediction_CE}(b)) and passed into the fine-tuned cross-encoder model (Step 2, \ref{fig:fine-tuning_prediction_CE}(b)), which generates generate a relevance score. This score, ranging from 0 to 1, estimates the likelihood of a method being buggy. This step refines the retrieved documents, enhancing the relevance between the bug report and source code (Step 4, Fig. \ref{fig:schematic-workflow-diagram}).

\subsection{Query Reformulation}
Although the above steps narrow down our search space, we further refine the relevance estimation between bug reports and source code segments by reformulating our search queries. In the following section, we discuss how we reformulate our queries by leveraging the LLM's reasoning capabilities as follows.

\begin{algorithm}[!ht]
\caption{Extract Keywords}
\label{alg:extract_keywords}
\begin{algorithmic}[1]

\Require Document $Doc$, number of top keywords $N$, trade-off parameter $\lambda$ ($0 \leq \lambda \leq 1$)

\Procedure{ExtractKeywords}{$Doc,N,\lambda$}
    \State $D \leftarrow \{\}$ \Comment{Initialize an empty dictionary}
    \State $T \leftarrow \textsc{Preprocess}(Doc)$ \Comment{Get tokens}
    \For{each $t \in T$}
        \State $E \leftarrow \text{Embed}(t)$
        \State $D[t] \leftarrow E$ \Comment{Add embedding for a token}
    \EndFor
    
    \State $B \leftarrow \text{Embed}(Doc)$ \Comment{Compute document embedding}
    \State $K \leftarrow \emptyset$ \Comment{Initialize selected keyword set}
    \State $C \leftarrow D.keys()$ \Comment{Candidate tokens}
    
    \While{$|K| < N$ and $C \neq \emptyset$}
        \State $Scores \leftarrow \{\}$
        \For{each $t \in C \setminus K$}
            \State $s_d \leftarrow \textsc{CosineSimilarity}(D[t], B)$ \Comment{Similarity with document}
            \If{$K \neq \emptyset$}
                \State $s_k \leftarrow \max\limits_{k \in K} \textsc{CosineSimilarity}(D[t], D[k])$
            \Else
                \State $s_k \leftarrow 0$ \Comment{No diversity penalty if no keywords selected}
            \EndIf
            \State $MMR \leftarrow \lambda \cdot s_d - (1 - \lambda) \cdot s_k$ \Comment{Compute MMR score}
            \State $Scores[t] \leftarrow MMR$
        \EndFor
        
        \State $t^* \leftarrow \arg\max\limits_{t \in C \setminus K} Scores[t]$ \Comment{Select token with highest MMR score}
        \State $K \leftarrow K \cup \{t^*\}$
        \State $C \leftarrow C \setminus \{t^*\}$ \Comment{Remove selected token from candidates}
    \EndWhile
    
    
    \State \textbf{return} $K$ \Comment{Return top-N keywords}
\EndProcedure
\end{algorithmic}
\end{algorithm}

 \subsubsection{Pre-training of Large Language Models for Software Bugs}
 We extract salient keywords from a bug report and relevant code segments (Steps 5a, 5b, Fig. \ref{fig:schematic-workflow-diagram}), using a Transformer-based model. 
 By leveraging the self-attention mechanism \cite{attention_is}, Transformer models can focus on the most relevant parts of the text, understand the context, and represent the text comprehensively using high-dimensional vectors. Their representations can be further adapted to different application domains through pre-training using domain specific unstructured data \cite{sun2020sifrank}. 

 \label{sec:pretraining_model}
To help capture salient keywords from bug reports and source code, we pre-trained a transformer model, CodeT5, using bug reports with masking. The process of collecting the pre-training dataset is explained in Section \ref{section:dataset_constraction}. We chose the CodeT5 model for embedding generation since it has been trained on both natural language texts and source code, which is ideal for bug reports containing various elements including text, code snippets, and program elements. Masked pre-training \cite{BERT} involves randomly masking a portion of the input tokens during training, encouraging the model to learn contextual representations by predicting the masked tokens. Bug reports contain valuable information about software issues, including descriptions of bugs, code snippets, and stack traces, etc.  
By incorporating the masking technique to learn from bug reports, we aimed to enhance CodeT5's general understanding of the domain-specific language, including its software engineering jargon and bug-related linguistic nuances. 

During the pre-training phase, we implemented a masking mechanism where 15\% of the tokens in each input sequence (i.e., bug reports) were randomly selected and substituted with a special \texttt{<extra\_id\_XX>} token. This approach aligns with the original T5 \cite{T5} model's convention, where XX serves as a unique identifier assigned to each masked token. The identifier follows a continuous numbering system, ensuring that each masked token in the input sequence receives a distinct number in sequential order (e.g., \texttt{<extra\_id\_0>, <extra\_id\_1>, <extra\_id\_2>}, and so forth). These input sequences with the masked tokens are then fed into the model to predict the original tokens based on contextual cues. By training the model to reconstruct the original tokens from the masked input, it acquires the ability to capture the general contextual understanding of bugs from bug reports.

 \subsubsection*{\textbf{Keywords from Bug Reports:}}  Keyword extraction involves selecting a subset of words or phrases that capture the main theme of a text document, which can support various subsequent tasks (e.g., Information Retrieval \cite{keyword_IR}). For extracting keywords from bug reports, we employ EmbedRank \cite{embedRank} due to its simplicity and compatibility with Transformer models using the KeyBERT \cite{keybert_library} library. 
 The Algorithm \ref{alg:extract_keywords} outlines the keyword extraction process.
We begin by pre-processing a bug report using standard text pre-processing techniques, such as removing stop words and punctuation marks. Next, we employ our pre-trained CodeT5 model to embed each token of the bug report. Similarly, we apply the technique to embed the entire bug report. Then we calculate cosine similarity \cite{cosine_similarity} to measure the semantic proximity of each token to the bug report. Based on these scores, we select the top-N most similar keywords (Step 5b, Fig. \ref{fig:schematic-workflow-diagram}) for the subsequent steps. To maximize diversity in the chosen keywords, we employ the Maximal Marginal Relevance (MMR) \cite{mmr_algorithm} algorithm and set the MMR parameter $\lambda = 0.5$, following the authors' recommendation.
Since the number of keywords affects retrieval, we also determined the optimal value of N through a controlled experiment, as discussed in $RQ_2$.

 \subsubsection*{\textbf{Keywords from Code Segments:}} We follow a similar approach to extract keywords from source code segments (Step 5a, Fig. \ref{fig:schematic-workflow-diagram}). However, instead of processing entire source documents, we consider only the relevant code segments identified by the cross-encoder with a confidence score above a predefined threshold (e.g., 0.5). If multiple code segments within a document are deemed relevant, we concatenate them into a single code block before applying Algorithm \ref{alg:extract_keywords} for keyword extraction. We detail the selection of this threshold through a controlled experiment in $RQ_3$.

 \subsubsection{Reformulating the Query:} 
  After extracting keywords from the bug report and relevant code segments, we leverage them to reformulate our queries and to enhance the retrieval (Step 6, Fig. \ref{fig:schematic-workflow-diagram}). We first measure the semantic similarity between the bug report keywords and the code keywords using cosine similarity, leveraging embeddings from the CodeT5 model. Our goal was to detect the top relevant documents and capture their overlapping tokens to enhance the keywords from the bug report. This reformulated query based on such an enhancement is then used in the subsequent steps.

\subsection{Bug Localization}
Once we have the reformulated query, we rerank the top-K results (e.g., 100) initially retrieved from the Elasticsearch (Step 7, Fig. \ref{fig:schematic-workflow-diagram}). Similar to the initial retrieval, we use the BM25 algorithm to rerank with the query constructed in the previous step. As a result, it provides a more refined set of final results (Step 8 of Fig. \ref{fig:schematic-workflow-diagram}) for bug localization. Our goal was to place the buggy documents at the top positions within the ranked list through the reranking. Then the developers will encounter the buggy documents earlier and spend less time analyzing the false-positives.

\section{Experiment}
\label{sec:experiments_iqloc}

We evaluate our approach using the Bench4BL benchmark dataset and three appropriate performance metrics. We conduct our experiments on a cluster computing system equipped with an NVIDIA GPU with 16 GB of vRAM and compare our technique, IQLoc, against eight baseline techniques from the literature. Using our experiments, we attempt to answer four research questions as follows:

\begin{itemize}

\item \textbf{\textit{RQ$_1$}}: How does IQLoc perform in bug localization in terms of evaluation metrics? 

\item \textbf{\textit{RQ$_2$}}: \textit{(a)} How does query length affect IQLoc's performance in bug localization? \textit{(b)} What is the rationale for choosing the CodeT5 model for reformulating query? \textit{(c)} Does the use of pre-trained models enhance the performance of these queries?

\item \textbf{\textit{RQ$_3$}}: How does the cross-encoder model perform in identifying relevant buggy source code based on program semantics?


\item \textbf{\textit{RQ$_4$}}: \textit{(a)} Can IQLoc outperform the existing baseline techniques in bug localization? \textit{(b)} How does it perform in localizing different types of bug reports compared to the baseline techniques?

\end{itemize}

\begin{table}[!ht]

    \caption{Bench4BL Dataset Summary}
    \vspace{-0.5\baselineskip} 
    \centering
    \begin{tabular}{|c|c|c|c|}
    \hline
    Project & Subject-systems & Bug reports \\
    \hline
    \hline
    Apache & 13  & 4,503 \\
    \hline
    Jbosss & 8  & 1,505 \\
    \hline
    Spring & 25  & 3,451 \\
    \hline
    Old Subject & 5  & 558 \\
    \hline
    \hline
    Total & 51 & 10,017 \\
    \hline
    \end{tabular}
    \label{table:bench4bl}
\end{table}





\subsection{Dataset Construction}
\label{section:dataset_constraction}

We use Bench4BL \cite{bench4bl}, a benchmark dataset, for our experiments. Table \ref{table:bench4bl} summarizes the Bench4BL dataset. Since the dataset is relatively old, we refine and extend it with more recent bug reports. Table \ref{table:bench4blExtended} shows our refined and expanded dataset. In total, we spent over 80 hours refining and expanding the dataset. The following sections detail our refinement process, the collection of new bug reports, and the dataset’s split for training and testing in bug localization. 

\subsubsection{Refinement of Bench4BL Dataset:}
\label{DS_expansion_train_test}
The Bench4BL \cite{bench4bl} dataset contains curated bug reports from various Java-based community projects, including Apache, JBoss, and Spring. It captures 10,017 bug reports from 51 subject systems along with their buggy/fixed versions of code and other relevant metadata. Table \ref{table:bench4bl} provides a summary of the benchmark dataset.

From our initial analysis of the benchmark dataset, we noticed that many projects or bug reports lacked appropriate versioning information, making them unsuitable for our experiment. We thus dropped the bug reports without any version details and ensured that the remaining reports could be traced back to both buggy and bug-free versions of the code. During this process, we encountered several challenges, including cases where either the buggy or fixed version was missing in the code repository, as well as instances where the mentioned ground-truth document did not exist. We discarded the bug reports with such discrepancies, ultimately retaining a dataset of 5,753 bug reports from the three large community projects (i.e., Apache, JBoss, and Spring).

\begin{figure*}
    \centering
    \includegraphics[width=0.9\linewidth]{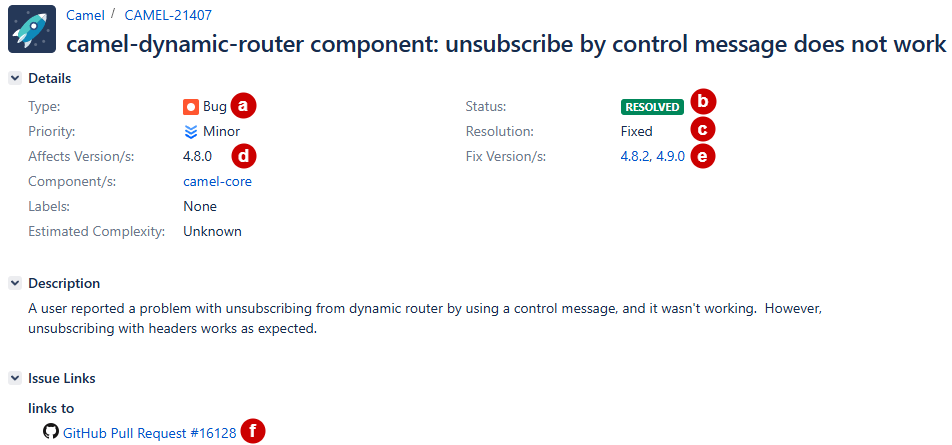}
    \caption{An Example Bug Report from JIRA}
    \label{fig:jira_bug_report}
\end{figure*}

\subsubsection{Expanding the Bench4BL Dataset:} After refining the original Bench4BL dataset, we expanded it by adding more recent bug reports for our experiments. In particular, we collected bug reports that were submitted by September 2024. Our expansion is limited to bug reports from the same projects and subject systems as the original dataset. These systems are managed by GitHub (e.g., Spring)  and JIRA (e.g., JBoss, Apache). To collect new bug reports, we employed two distinct approaches tailored to these issue-tracking systems.

To collect bug reports from JIRA-based issue tracking systems, we used the JIRA API \cite{JiraAPI}. We also filtered issues that were of type \textcircled{a} ‘Bug,’ had a status of \textcircled{b}‘Resolved,’ a resolution of \textcircled{c} ‘Fixed,’ and included both \textcircled{d} a buggy version and \textcircled{e} a fixed version (Fig. \ref{fig:jira_bug_report}). Additionally, we considered bug reports containing explicit \textcircled{f} Git pull request links to ensure a proper mapping between the buggy and fixed versions, reducing false positives. Once we collected the bug reports from JIRA, we used the GitHub API to extract the corresponding buggy and fixed files from the linked pull requests. Since a pull request can contain multiple commits, we considered only accepted and verified commits. When multiple commits were associated with a pull request, we included all files across these commits as a part of the fix for that bug report. This approach accounts for bug fixes that evolve over time, often involving multiple developers addressing different aspects of the same bug. Additionally, it captures cases where bug reports were closed and later reopened for further fixes. By including all relevant files, we ensure a comprehensive representation of the bug-fix process. Finally, we ensured that the versions specified in JIRA issues (e.g., semantic version, 4.8.2) matched the version tags in GitHub (e.g., camel-4.8.2) using regex-based validation, as the version strings in JIRA could differ from those in GitHub.

\begin{figure*}
    \centering
    \includegraphics[width=0.9\linewidth]{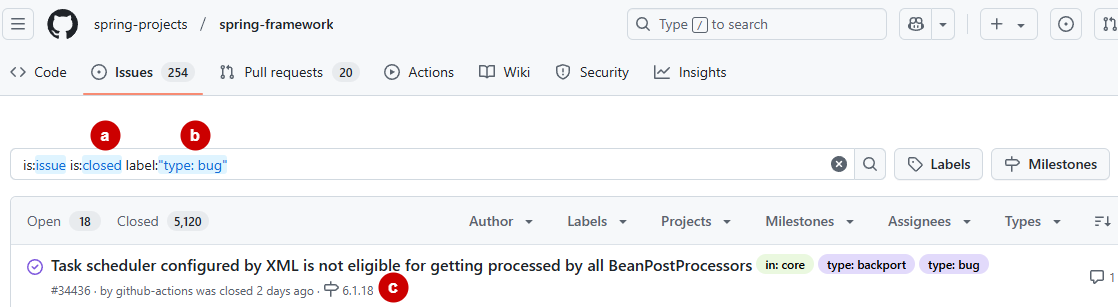}
    \caption{GitHub Issue Selection}
    \label{fig:git_issue_selection}
\end{figure*}

\begin{table}[ht]

    \caption{Refined and Expanded Dataset}
    \vspace{-0.5\baselineskip} 
    \centering
    \begin{tabular}{|c|c|c|c|}
    \hline
    Project & Subject-systems & Major versions & Bug reports \\
    \hline
    \hline
    Apache & 11 & 428 & 3,145 \\
    \hline
    Jbosss & 6 & 303 & 1,416 \\
    \hline
    Spring & 25 & 847 & 2,922 \\
    \hline
    \hline
    Total & 42 & 1,578 & 7,483 \\
    \hline
    \end{tabular}
    \label{table:bench4blExtended}
\end{table}

\begin{figure}[h]
    \centering
    \includegraphics[width=0.6\linewidth]{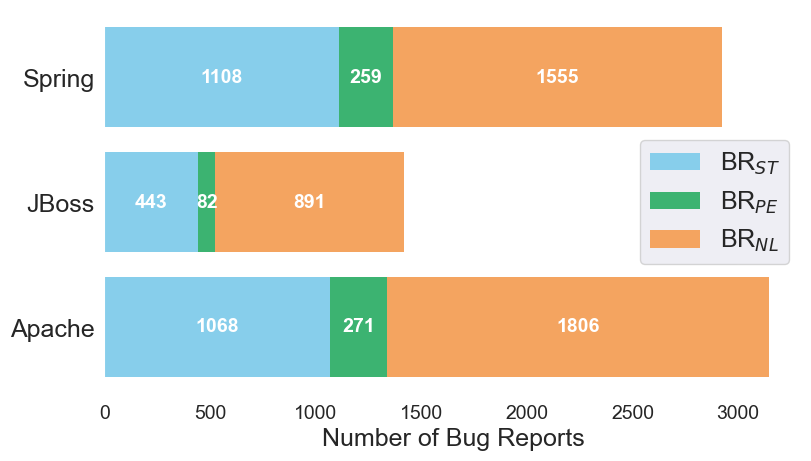}
    \caption{Classification of Bug Reports}
    \label{fig:dataset_classification}
\end{figure}

On the other hand, when collecting bug reports from GitHub, we considered issues that are \textcircled{a} 'closed' and labeled as \textcircled{b} 'type: bug' (Fig. \ref{fig:git_issue_selection}). To determine the version in which a bug was fixed and reduce false positives, we collected bug reports that had explicit \textcircled{c} milestones attached, indicating the version(s) in which the bug was resolved. Since GitHub issue reports do not explicitly define the buggy version, we identified the immediate previous version tag associated with the first bug-fix commit associated with the issue in the branch tree. While this may not always correspond to the exact commit or version where the bug was introduced, it allows us to determine the latest version in which the bug exists. After that, for resolving the buggy files, we applied the same approach as used for JIRA bug reports.

Once we collected the bug reports and resolved the buggy files, we dropped the bug reports related to configuration bugs (i.e., IML, XML). Following this procedure, we complemented the original Bench4BL dataset with 1,730 recent bug reports. In total we curated 7,485 bug reports from 42 subject-systems across 1,578 buggy/fixed versions. 

To gain further insight, we classify bug reports in our experimental dataset based on their content, as was done by existing literature \cite{blizzard}:

\begin{itemize}

    \item BR\textsubscript{ST} (Bug Reports with Stack Traces): Includes stack traces along with text or program elements. Queries from these reports are generally noisy.

    \item BR\textsubscript{PE} (Bug Reports with Program Elements): Contains program elements (e.g., method invocations, package names) but no stack traces. Queries from these reports are considered rich.

    \item BR\textsubscript{NL} (Bug Reports with Natural Language Only): Lacks both program elements and stack traces. Queries from these reports are generally less informative.

\end{itemize}

To classify them, we use regular expressions adapted from the work of Rahman et al. \cite{blizzard}. Fig. \ref{fig:dataset_classification} shows the classification of our curated dataset.

\begin{table}[!t]

    \caption{Train, Validation and Test-sets}
    \vspace{-0.5\baselineskip} 
    \centering
    \begin{tabular}{|l|l|l|l|l|}
    \hline
        Split Type  & Training & Validation & Test & Total \\ \hline \hline
        Random Split & 5,238 & 748 & 1,497 & 7,483  \\ \hline
        Time-wise Split & 5,236 & 746 & 1,501 & 7,483  \\ \hline
    \end{tabular}
    \label{table:bench4bl_refined}
\end{table} 

\begin{figure}[h]
    \centering
    \begin{tabular}{cc} 
        \includegraphics[width=0.48\linewidth]{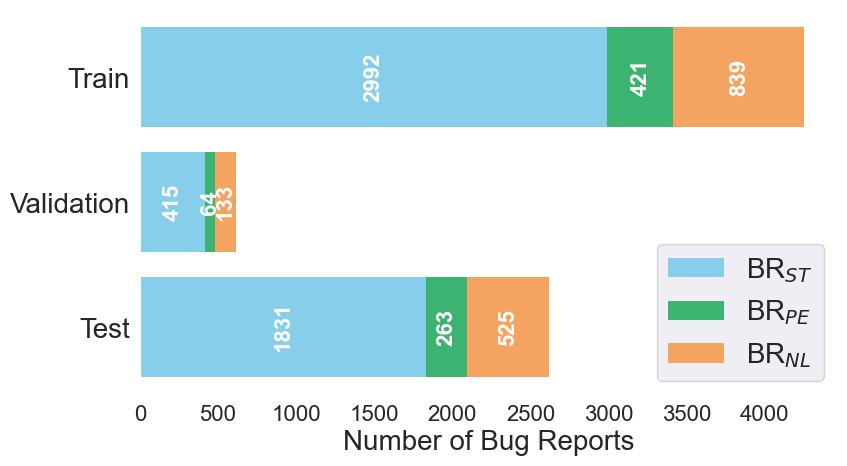} &
        \includegraphics[width=0.48\linewidth]{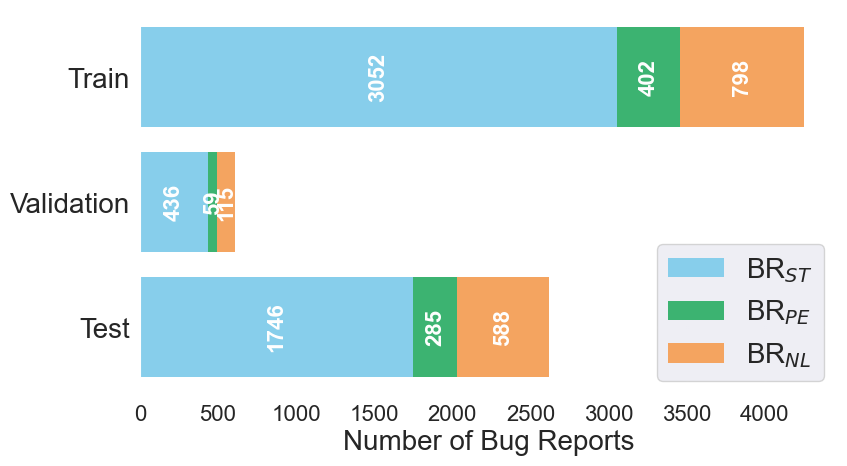} \\

        \small (a) Random Split & \small (b) Time-wise Split
    \end{tabular}
    \caption{Distribution of Bug Reports in Different Dataset-Splits}
    \label{fig:classification_train_test_valid}
\end{figure}

\subsubsection{Train and Test Set:}
For our experiments, we split the dataset into training, validation, and test sets using a 70:10:20 ratio, following the standard machine learning practice for data partitioning \cite{dataset_splits, data_splitting}. In our dataset splitting, we adopted two strategies, as was done by earlier studies \cite{time_split, random_split}. The first approach utilized random selection with shuffling, where the dataset was randomly divided into training and test sets. To ensure an unbiased random split, we conducted this process five times, generating five randomly shuffled datasets to facilitate separate experiments. This process delivered 5,238 bug reports for training, 1,497 for testing and 748 for validation on random trials in each of the five sets. The second strategy was a time-wise split. This type of splitting divides data into training and test sets based on their chronological order, simulating real-world scenarios where models are trained on past data and evaluated on future instances. For each subject system, we sorted the bug reports by their submission dates, split them individually into train, validation and test sets, and then combined them with similar splits from other systems. This process resulted in 5,236 bug reports for training, 1,501 for testing and 746 bug reports for validation. Table \ref{table:bench4bl_refined} summarizes our curated dataset for the experiments. Also, Fig. \ref{fig:classification_train_test_valid} shows how classified bug reports (i.e., BR\textsubscript{ST}, BR\textsubscript{PE}, BR\textsubscript{NL}) are distributed across our training, validation and test dataset. Note, for random-split, the distribution shows the average of five random-splits.

\begin{table}[!t]
    \caption{Cross-Encoder Dataset (Random Split)}
    \centering
    \begin{tabular}{|l|l|l|l|l|}
    \hline
    Dataset Type & Bug Reports & $q, d^+$ Pairs (Avg.) & $q, d^-$ Pairs (Avg.) & Total (Avg.) \\ \hline\hline
    Training   & 5,238  & 12,694 & 50,776 & 63,470 \\ \hline
    Test       & 1,497  & 3,715  & 14,860 & 18,575 \\ \hline
    Validation & 748    & 1,829  & 7,316  & 9,145  \\ \hline
    \end{tabular}
    \label{tab:crossencoder_random}
\end{table}

\begin{table}[!t]
    \caption{Cross-Encoder Dataset (Time-wise Split)}
    \centering
    \begin{tabular}{|l|l|l|l|l|}
    \hline
    Dataset Type & Bug Reports & $q, d^+$ Pairs & $q, d^-$ Pairs & Total \\ \hline\hline
    Training   & 5,236  & 12,727 & 50,908 & 63,635 \\ \hline
    Test       & 1,501  & 3,820  & 15,280 & 19,100 \\ \hline
    Validation & 746    & 1,691  & 6,764  & 8,455  \\ \hline
    \end{tabular}
    \label{tab:crossencoder_timewise}
\end{table}

\subsubsection{Expanding Training Set for Cross-Encoder:}
To fine-tune our cross-encoder model, we extract both the buggy and bug-free versions of the code associated with each bug report from the GitHub repository of the respective subject system. Using a Diff tool \cite{pypiDifftool}, we compare between buggy and bug-free code and identify the method bodies in the buggy document that were changed to correct the bug. These altered method bodies are selected as positive instances $(q, d^+)$ for our model training against each bug report, with a label of 1 assigned to them. This process is illustrated in Fig. \ref{fig:fine-tuning_prediction_CE}.

In our dataset, we observed that each bug resulted in changes to a minimum of 1 and a maximum of 7 source code documents. To construct the training set for our cross-encoder, which determines the semantic relevance between a bug report and a document, we process all documents individually and generate multiple instances of positive associations. For example, if a bug report requires changes to three documents, we create three training instances, linking the altered method bodies to the bug report as positive instances for training. Our approach ensures that each training instance has a limited number of tokens, ensuring compatibility with the transformer model, which has a token limit of 512.

To generate negative samples $(q, d^-)$, we randomly select method bodies associated with different bug reports for a given report at hand. We only consider source code from different subject systems for these negative cases. This process is reiterated four times to yield sufficient training instances for the model, and they are labeled as 0. The choice of four negative instances against each positive instance is based on a study conducted by Huang et al. \cite{sem_srch_pos_neg_sample}. They found no notable distinctions across different numbers of negative samples and also endorsed the 4:1 ratio.

For the validation and test data, we followed the same approach to generate positive and negative samples. Tables \ref{tab:crossencoder_random} and Table \ref{tab:crossencoder_timewise} summarize the training, validation, and test sets for the cross-encoder across two types of experiments.


\subsubsection{Dataset for Pre-training Model with Bug Reports:}
To pre-train an existing baseline language model (e.g., CodeT5) with domain-specific data, we collected thousands of bug reports from GitHub repositories hosting Java-based projects using the GitHub API. Our selection process involved choosing the top 100 repositories based on their star count and thus ensuring that they remained active up to the date of selection. To maintain data integrity, we targeted the issue reports labeled as 'bugs' and collected reports submitted before April 2024. 
Additionally, we confirmed that none of these repositories were already included in the Bench4BL dataset to prevent any bias. After collecting the bug reports, we meticulously cleaned them to retain only their textual content using Beautiful Soup \cite{BeautifulSoupDocs} and excluded repositories with fewer than five bug reports. Finally, all these steps resulted in a dataset of 70,884 bug reports from 74 repositories. Then, these bug reports were used to pre-train the CodeT5 model.

%


\subsection{Evaluation Metrics}
To evaluate IQLoc in bug localization, we use three widely used metrics- MAP, MRR, and HIT@K. These metrics have been frequently used by the relevant literature on IR-based bug localization \cite{blizzard, rahman2021forgotten}, and thus are highly relevant to our approach. To perform ablation study involving our cross-encoder model, we used four commonly used metrics -- Accuracy, Precision, Recall, and F1 score.

\subsubsection*{\textbf{Mean Average Precision (MAP)}}

Precision@K refers to precision for each occurrence of the buggy source document in the ranked list. Average Precision calculates the average precision@K for all buggy documents against a search query. Therefore, Mean Average Precision (MAP) is computed by averaging the AP values across all queries (Q) from a dataset.

\[ 
P_k = \frac{\textit{No. of Relevant Items in Top-}k}{k} 
\]
\[
AP@K = \frac{1}{|D|} \sum_{k=1}^{K} P_k \times B_k
\]
\[
MAP = \frac{1}{|Q|} {\sum_{q=1}^{Q} AP@K_q}
\]

Here, $P_k$ calculates the precision for the $k^{th}$ element of the top $K$ items in the ranked list returned by a query. $AP@K$ computes the average precision for a list of $K$ results against a query, utilizing $B_k$ to determine if the $k^{th}$ document is buggy or not. $B_k$ outputs 1 for a match with the ground truth and 0 otherwise. $D$ represents the set of relevant instances that match the ground truth documents against a query. Finally, $MAP$ calculates the mean of the average precision ($AP@K$) across all the individual queries $q$ in the set $Q$.

\subsubsection*{\textbf{Mean Reciprocal Rank (MRR)}}
Reciprocal Rank (RR) is associated to the rank of the first relevant result retrieved by a technique. It calculates the reciprocal of the rank of the first relevant source document within the ranked list returned by each query.
\[ RR_q = \frac{1}{\textit{Rank of First Relevant Item}} \]

\[
MRR = \frac{1}{|Q|} \sum_{q=1}^{|Q|} RR_q
\]

Here, $RR_q$ represents the Reciprocal Rank for a specific query $q$. Thus, the Mean Reciprocal Rank ($MRR$) is calculated as the mean of the Reciprocal Ranks ($RR_q$) for all individual queries $q$ in the set $Q$.




\subsubsection*{\textbf{HIT@K}}
HIT@K or Recall@Top-K \cite{saha_bleuir} refers to the proportion of queries for which a technique returns at least one relevant document among the top $K$ retrieved results. The higher the HIT@K values, the better the performance of a bug localization technique.

\[
HIT@K = \frac{1}{|\mathcal{Q}|} \sum_{q=1}^{|Q|} \begin{cases}1, & r_q \in \mathcal{G} \\ 0, & \text{otherwise}\end{cases}
\]

Here, $Q$ is the set of all queries and $r_q$ is a binary indicator function that returns 1 if the query $q$ has at least one ground truth item $r_q \in G$ within the top-K results, and 0 otherwise.

\subsubsection*{\textbf{Accuracy}}
Accuracy is a metric that determines the proportion of correctly predicted instances out of the total instances evaluated.
\[
\text{Accuracy} = \frac{TP + TN}{TP + FP+ TN + FN}
\]

Here, $TP$ (True Positive) and $TN$ (True Negative) represent instances that are \textit{correctly predicted} as positive and negative, respectively. Conversely, $FP$ (False Positive) and $FN$ (False Negative) represent instances that are \textit{incorrectly predicted} as positive and negative, while in reality, they are negative and positive instances, respectively. In the context of bug localization, positive instances denote buggy documents, while negative instances denote non-buggy ones.

\subsubsection*{\textbf{Precision}}
Precision measures the proportion of correctly \textit{predicted} positive instances out of all positive \textit{predictions} made by a model. 

\[
\text{Precision} = \frac{TP}{TP + FP}
\]

\subsubsection*{\textbf{Recall}}
Recall, also known as sensitivity, measures the proportion of correctly \textit{predicted} positive instances out of all \textit{actual} positive instances in the dataset. 

\[
\text{Recall} = \frac{TP}{TP + FN}
\]

\subsubsection*{\textbf{F1 Score}}
The F1 Score provides a single measure that balances between precision and recall. It is calculated as the harmonic mean of precision and recall. This balanced measure is particularly valuable in scenarios where false positives and false negatives have different implications, ensuring a robust assessment of a model's performance.

\[
\text{F1 Score} = 2 \times \frac{Precision \times Recall}{Precision + Recall}
\]

\begin{table*}[!ht]
    \caption{Performance of IQLoc}
    \vspace{-0.5\baselineskip} 
    \centering
    \begin{tabular}{|l|l|l|l|l|l|}
    \hline
         Split Type & MAP & MRR & HIT@1 & HIT@5 & HIT@10  ~ \\ \hline\hline
        Random Split & 0.493 & 0.520 & 0.423	& 0.647	& 0.721 \\ \hline
        Time-wise Split & 0.520 & 0.553 & 0.466	& 0.669	& 0.735 \\ \hline

    \end{tabular}
    \label{tab:rq1_performance}
\end{table*}

\begin{table}[!ht]
    \caption{Impact of the Selection of Top-K Results from Elasticsearch}
    \centering
    \begin{tabular}{|l|l|l|l|l|l|l|l|l|l|}
    \hline
        Split Type & HIT@1 & HIT@5 & HIT@10 & HIT@50 & HIT@100 & HIT@200 ~ \\ \hline\hline
        Random Split & 0.389 & 0.619 & 0.703 & 0.843 & 0.889 & 0.929  \\ \hline
        Time-wise Split & 0.396 & 0.618 & 0.822 & 0.904 & 0.937 & 0.961  \\ \hline
    \end{tabular}
    \label{tab:baseline_vsm_br}
\end{table}




\subsection{Evaluting IQLoc}
\subsubsection{Answering RQ$_1$: Performance of IQLoc}
Table \ref{tab:rq1_performance} presents the performance of IQLoc in terms of Mean Average Precision (MAP), Mean Reciprocal Rank (MRR), and accuracy at different top-K results (HIT@1, HIT@5, HIT@10). For the randomly split dataset, we repeated our experiments five times with different random seeds, and report the mean performance across these runs. This repeated experiment helps reduce bias from any single random split of the dataset. On the other hand, we run our experiment once for the time-wise split dataset.

In the random split, IQLoc achieves a MAP of 0.493, indicating that, on average, the relevant documents  (a.k.a., buggy source code) rank higher than the irrelevant ones. Similarly, the MRR is 0.520, suggesting that the first relevant document is, on average, found within the top 2 positions. IQLoc achieves a HIT@1 of 0.423, indicating that 42.3\% of the bug reports have their relevant documents (a.k.a., buggy source code) retrieved as the top-ranked result. It also achieves a HIT@5 of 0.647, indicating that $\approx$65\% of the bug reports have returned at least one buggy document within the top 5 positions, whereas the HIT@10 measure is 0.721.

In the time-wise split, IQLoc demonstrates higher performance, achieving a MAP of 0.520 and an MRR of 0.553. Besides, a HIT@1 of 0.466 indicates a substantial performance improvement over its counterpart above. However, the improvement is more noticeable for HIT@5 and HIT@10 metrics, where nearly 67\% of the bug reports that have at least one buggy document retrieved within the top 5 positions and 73.5\% within the top 10 positions, respectively. These metrics suggest that IQLoc consistently performs well across different data splits. Moreover, its higher performance across all metrics in the time-wise split implies that IQLoc potentially captures temporal trends from past bug reports (i.e., time-wise split) and source code versions to identify recent bugs within the code. 

In our approach, we capture the top 100 results retrieved by the Elasticsearch module for subsequent reranking (Step 2, Fig. \ref{fig:schematic-workflow-diagram}). Previous studies in information retrieval have also used a subset of results to rerank \cite{bugcatcher_multilevel, CE_example_ranking1, CE_example_ranking2}. Our decision was made after carefully analyzing different top-K results. As demonstrated in Table \ref{tab:baseline_vsm_br}, for the random split test set, we observed a HIT@100 of 0.889 and a HIT@200 of 0.929. That is, for $\approx$89\% of the bug reports, a relevant result can be found within the top 100 search results, and for $\approx$93\% of the reports within the top 200 search results. This is a slight 4.5\% increase by considering an additional 100 results. For the time-wise split, this difference is even smaller, only 2.6\%. Besides, our reranking step relies on a Transformer-based cross-encoder model, which demands significant computing power. By considering only top 100 results, we thus strike a balance between the relevance of the results and the management of our computational resources.

\begin{table}[!ht]
    \centering
    \caption{Performance of IQLoc for Different Classes of Bug Reports}
    
    \caption*{(a) Time-wise Split}
    \begin{tabular}{l|c|c|c|c|c}
        \hline
        Model & MAP & MRR & HIT@1 & HIT@5 & HIT@10 \\
        \hline
        \hline
        \multicolumn{6}{c}{ST} \\
        \hline
        Baseline Elasticsearch     & 0.488 & 0.513 & 0.443 & 0.594 & 0.686 \\
        IQLoc  & 0.565 & 0.601 & 0.535 & 0.698 & 0.730 \\
        \hline
        \multicolumn{6}{c}{PE} \\
        \hline
        Baseline Elasticsearch      & 0.599 & 0.621 & 0.505 & 0.768 & 0.874 \\
        IQLoc  & 0.689 & 0.737 & 0.674 & 0.811 & 0.863 \\
        \hline
        \multicolumn{6}{c}{NL} \\
        \hline
        Baseline Elasticsearch      & 0.442 & 0.467 & 0.350 & 0.612 & 0.709 \\
        IQLoc  & 0.460 & 0.487 & 0.382 & 0.626 & 0.720 \\
        \hline
    \end{tabular}

    \vspace{0.5cm} 

    \caption*{(b) Random Split}
    \begin{tabular}{l|c|c|c|c|c}
        \hline
        Model & MAP & MRR & HIT@1 & HIT@5 & HIT@10 \\
        \hline
        \hline
        \multicolumn{6}{c}{ST} \\
        \hline
        Baseline Elasticsearch      & 0.492 & 0.513 & 0.421 & 0.637 & 0.721 \\
        IQLoc  & 0.561 & 0.592 & 0.503 & 0.708 & 0.755 \\
        \hline
        \multicolumn{6}{c}{PE} \\
        \hline
        Baseline Elasticsearch      & 0.582 & 0.632 & 0.542 & 0.759 & 0.819 \\
        IQLoc  & 0.655 & 0.687 & 0.602 & 0.771 & 0.843 \\
        \hline
        \multicolumn{6}{c}{NL} \\
        \hline
        Baseline Elasticsearch      & 0.423 & 0.454 & 0.353 & 0.592 & 0.680 \\
        IQLoc  & 0.432 & 0.456 & 0.350 & 0.596 & 0.686 \\
        \hline
    \end{tabular}
    \label{tab:performance_bug_classification}
\end{table}

 We also evaluate IQLoc’s performance in localizing different types of bug reports: bug reports containing Stack Trace (ST), Program Element (PE), and Natural Language (NL), as discussed in Section \ref{section:dataset_constraction}. In this case, we consider the Elasticsearch as a traditional baseline adapted from Apache Lucene \cite{lucene}. Table \ref{tab:performance_bug_classification} presents the results. In the time-wise split test set (Table \ref{tab:performance_bug_classification}(a)), IQLoc outperforms Elasticsearch by 15.77\% and 17.15\% in MAP and MRR, respectively for bug reports containing stack traces (ST). 
 It also improves such localization by detecting at least one buggy document in the top-10 positions, with gains of 6.41\%–20\% in HIT@1, HIT@5, and HIT@10. For bug reports with program elements (PE), performance improvements range from 5.6\% to 33.46\% in the time-wise split, with a 1.2\% drop in HIT@10. In contrast, our techniques improves in all metrics for the bug reports containing only natural language (NL), but the gains are smaller. IQLoc achieves only a 4.07\% increase in MAP and 4.28\% in MRR. HIT@K improvements range from 1.57\% to 9.14\%, which is lower than the gains for ST and PE. For the random split dataset, we observe a similar trend in localizing bug reports containing stack traces (ST) and program elements (PE). For ST, IQLoc outperforms Elasticsearch by 4.72\%–14.02\% across all metrics. For PE, the improvement ranges from 1.58\% to 12.54\%. However, similar to the time-wise split, performance gains for NL bug reports remain minimal, ranging from 0.44\% to 2.13\% across all metrics except HIT@1.
 
\begin{figure}[!t]
    \centering
    \begin{tabular}{cc} 
        \includegraphics[width=0.35\linewidth]{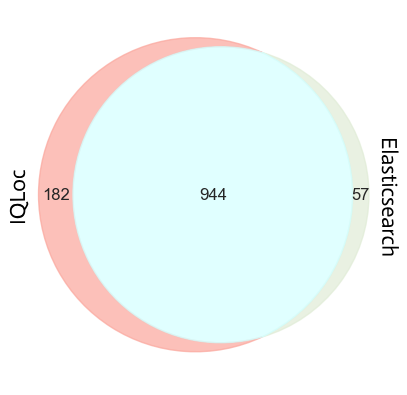}
        &
        \hspace{0.4cm}
        \includegraphics[width=0.35\linewidth]{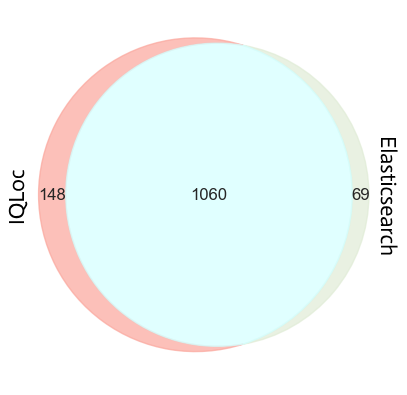}\\
        \small (a) Random Split & \hspace{0.4cm} \small (b) Time-wise Split
    \end{tabular}
    \caption{Impact of Query Reduction on Retrieval Performance}
    \label{fig:rq1_venn_dig}
\end{figure}

We also demonstrate how our reduced queries improve bug localization through document reranking in IQLoc. To illustrate this, we present two analyses from two different perspectives. Figure~\ref{fig:rq1_venn_dig} highlights the benefit of incorporating our reformulated queries during the reranking step by comparing two methods: baseline Elasticsearch and IQLoc, both evaluated using the top-10 results. In our analysis of randomly split test sets, we found that both techniques retrieved buggy source documents for 944 bug reports. However, IQLoc localizes at least one buggy document for 182 more bug reports that Elasticsearch could not. A similar pattern can be observed in the time-wise split dataset, where IQLoc localized 148 more bugs for which Elasticsearch could not succeed. It should be noted that baseline Elasticsearch also localized some unique bugs, but they were much lower in number.

\begin{figure}[!ht]
    \centering
    \begin{tabular}{c} 
        \includegraphics[width=0.7\linewidth]{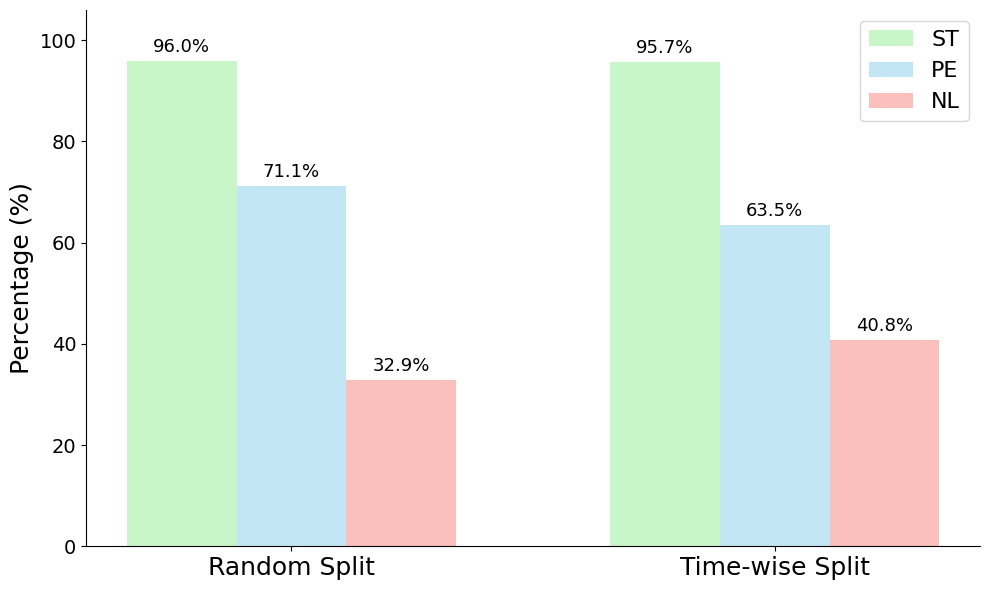}
    \end{tabular}
    \caption{IQLoc's Performance for Different Types of Bug Reports}
    \label{fig:rq1_class_low_quality}
\end{figure}

These analysis above led us to investigate the types of bug reports for which IQLoc might excel or fail during localization. Figure \ref{fig:rq1_class_low_quality} presents our findings for both time-wise and random splits. In the randomly split test dataset, IQLoc successfully localized the majority of bug reports containing stack traces (ST), accounting for 96.0\% of the localized cases, compared to 71.1\% for those containing program elements (PE) and 32.9\% for bug reports classified as natural language (NL). A similar pattern is observed in the time-wise split, where 95.7\% of bug reports with stack traces were localized, followed by 63.5\% with PE and 40.8\% with NL.

\begin{tcolorbox}[colback=lightergray,colframe=black,arc=1mm,boxrule=0.5pt]
\textbf{RQ1 Summary:} IQLoc demonstrates promising performance in localizing bugs, achieving a MAP score of up to 0.520— a 10.43\% improvement over baseline Elasticsearch. This improvement is driven by our query reduction strategy, which uses Transformer-based reasoning of the buggy code to better match between a query and the code. Additionally, IQLoc excels at handling various types of bug reports, successfully localizing up to 96.0\% of those containing stack traces.
\end{tcolorbox}


\begin{figure*}[!ht]
    \centering
    \begin{tabular}{cc} 
        \includegraphics[width=0.5\linewidth]{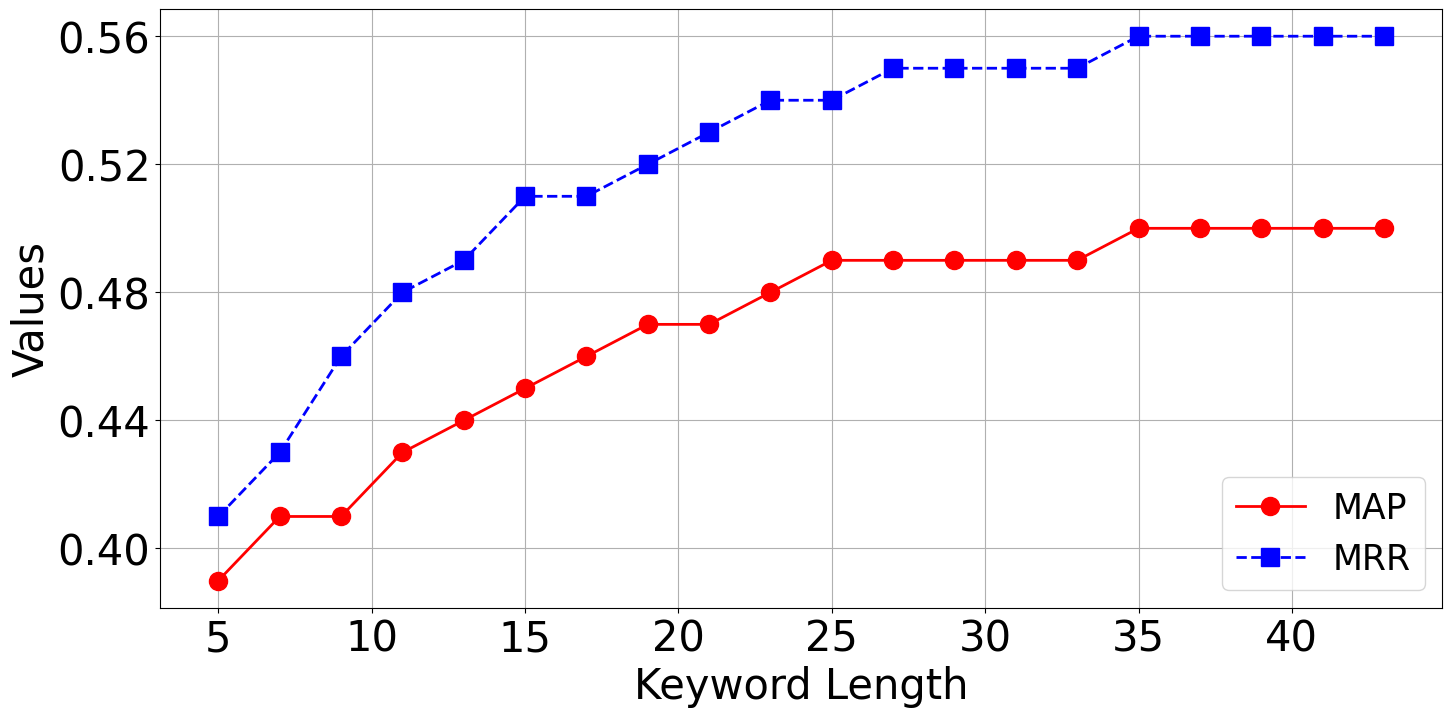} &
        \includegraphics[width=0.5\linewidth]{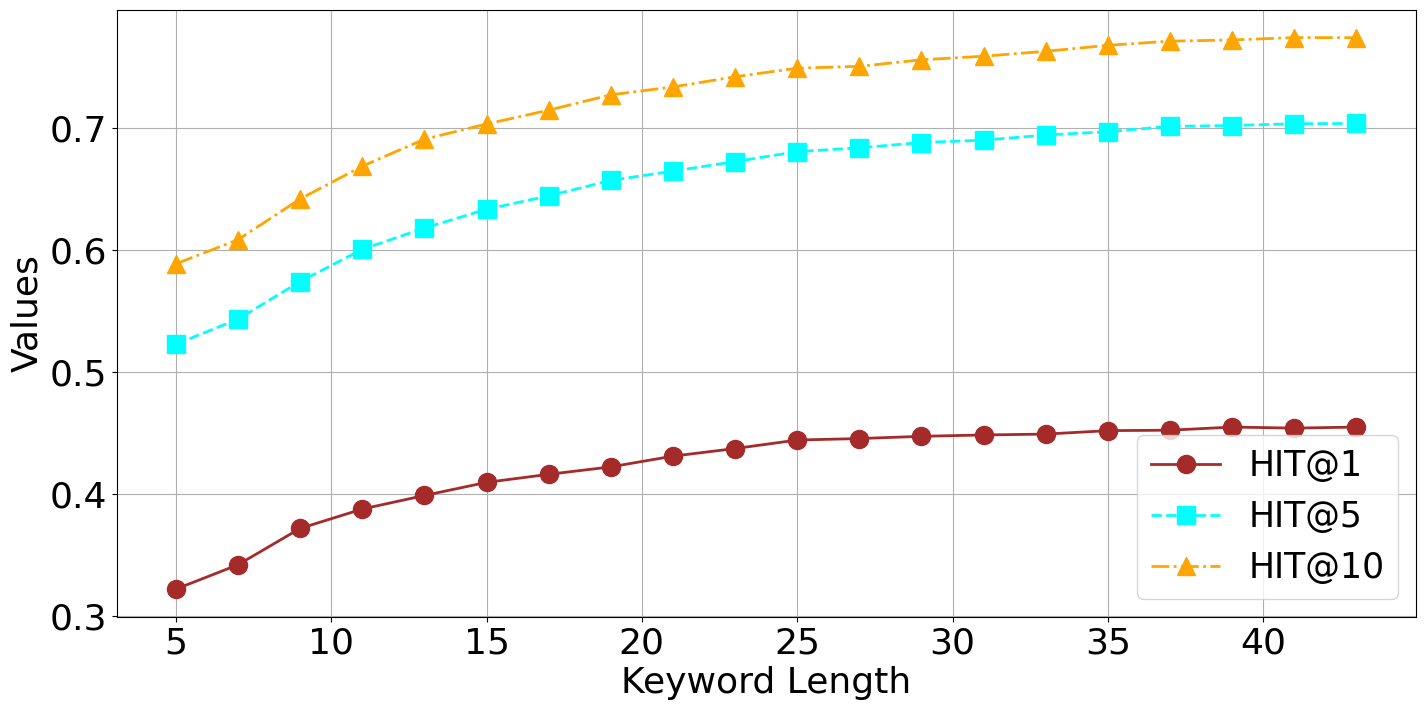}\\
        \small (a) MAP \& MRR & (b) HIT@K\\
    \end{tabular}
    \caption{Impact of Query Length on Bug Localization}
    \label{fig:keyword_length_best}
\end{figure*}

\begin{figure*}[!ht]
    \centering
    \begin{tabular}{cc} 
        \includegraphics[width=0.5\linewidth]{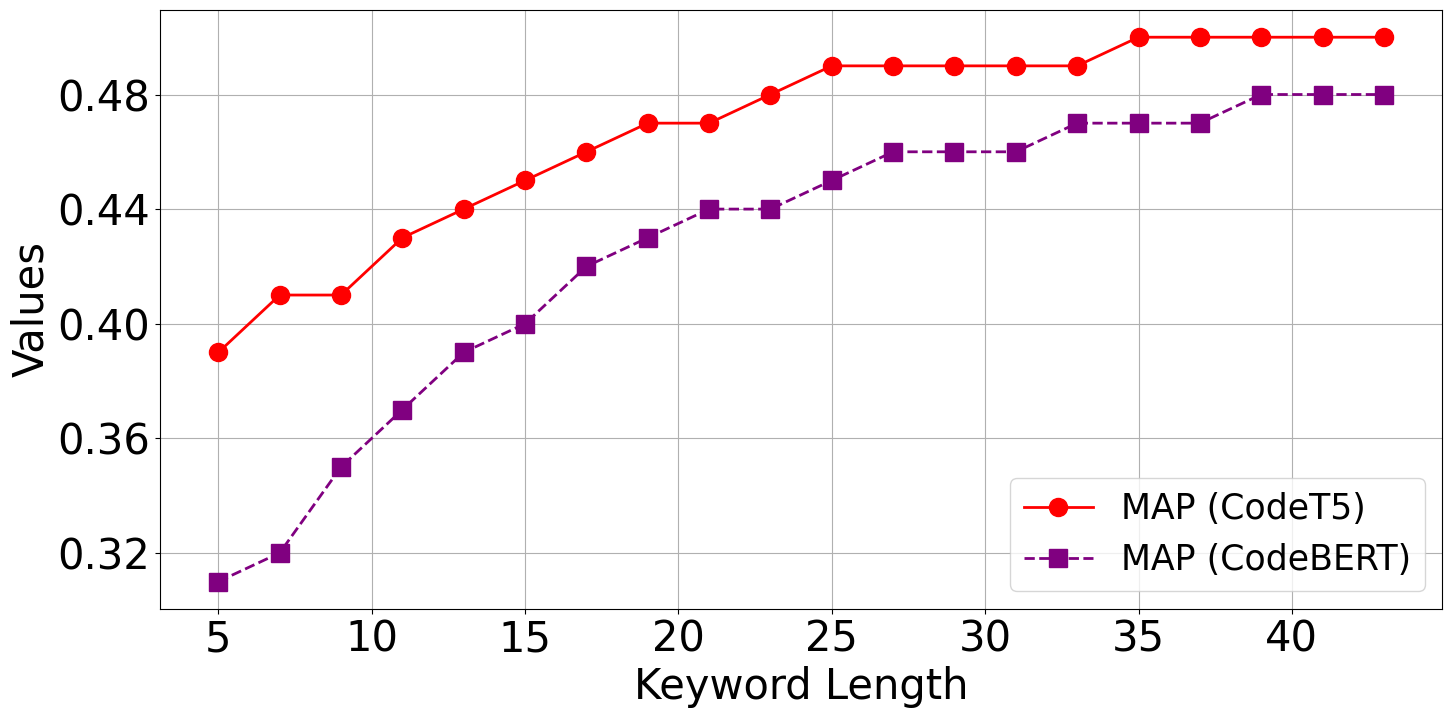} &
        \includegraphics[width=0.5\linewidth]{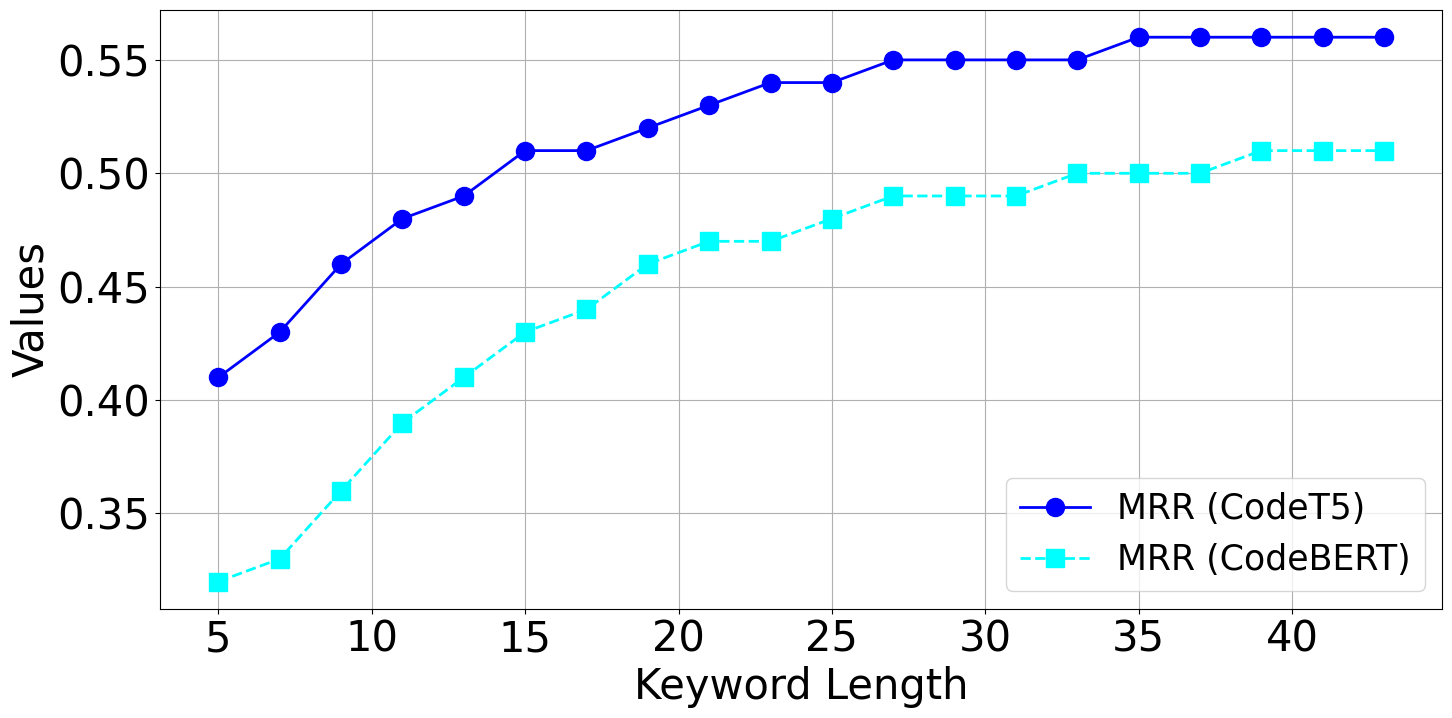}\\
        \small (a) MAP : CodeT5 vs. CodeBERT & (b) MRR : CodeT5 vs. CodeBERT\\
    \end{tabular}
    
    \caption{Choice of Pre-trained Models for Query Reformulation}
    \label{fig:codet5_vs_codebert}
\end{figure*}

\subsubsection{Answering RQ$_2$: Impact of Query Length and Embedding Models on IQLoc's Performance}

Fig. \ref{fig:keyword_length_best} shows how the length of our search queries (N) influences the performance of IQLoc. Selection of search keywords is a crucial aspect of our technique since the keywords can narrow down the search scope and position the relevant documents at higher ranks during bug localization. However, determining the optimal number of keywords in a query without compromising bug localization performance poses a challenge.

To determine an optimal length for search queries (N), we use a fixed pre-trained model, the baseline CodeT5 \cite{wang2021codet5} model, during document reranking step while varying the length of queries. From Fig. \ref{fig:keyword_length_best} we see that the HIT@1 increases from 0.32 with 5 keywords to a maximum of 0.455 with a keyword length of 43, representing a $\approx$42\% increase. Such a trend is consistent across other metrics, with improvements of 27.64\%, 37.61\%, 34.61\%, and 31.52\% for MAP, MRR, HIT@5, and HIT@10, respectively. However, we observe diminishing returns in performance improvement for HIT@K as query length increases, reaching a plateau around the length of 15, where performance improvement slows down. To balance between performance and query specificity, we thus chose a maximum query length of 15 for our experiments.

\begin{figure*}[!ht]
    \centering
    \small
    \begin{tabular}{cc} 
        \includegraphics[width=0.5\linewidth]{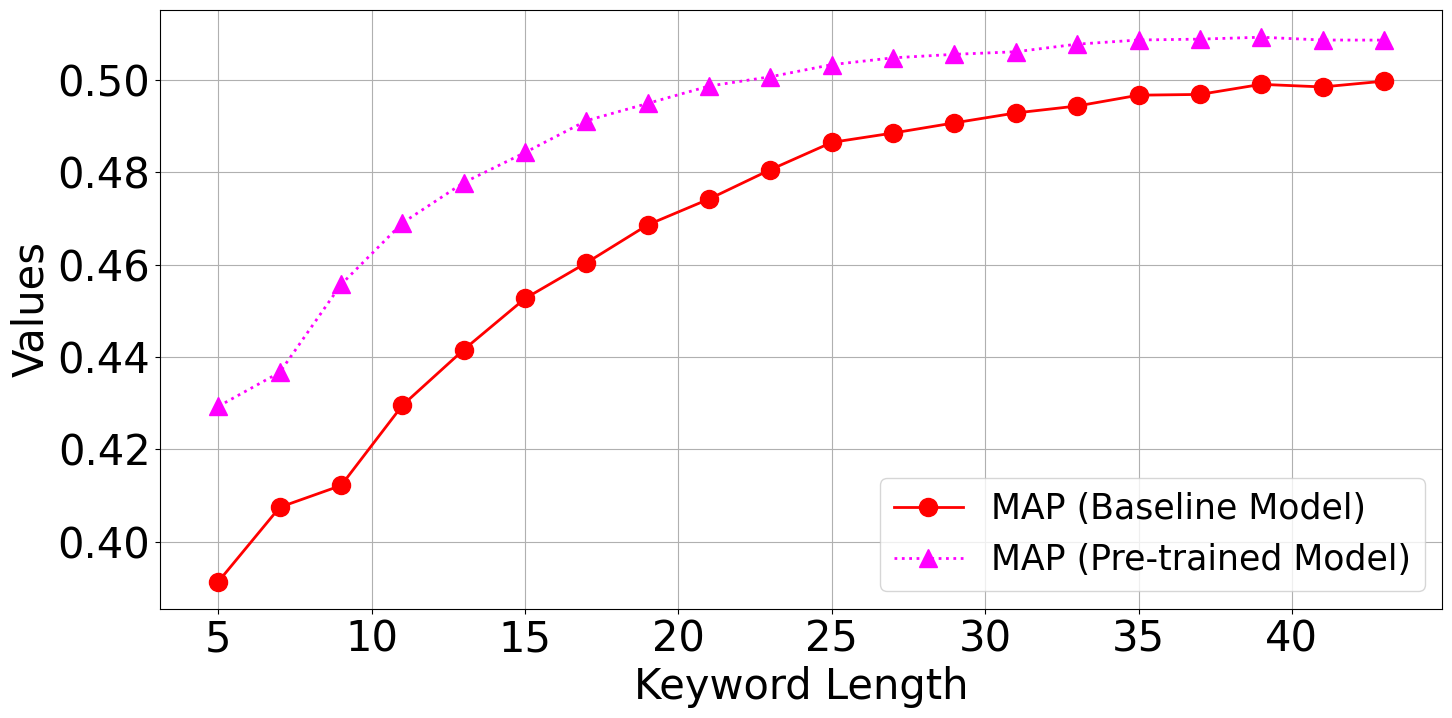} &
        \includegraphics[width=0.5\linewidth]{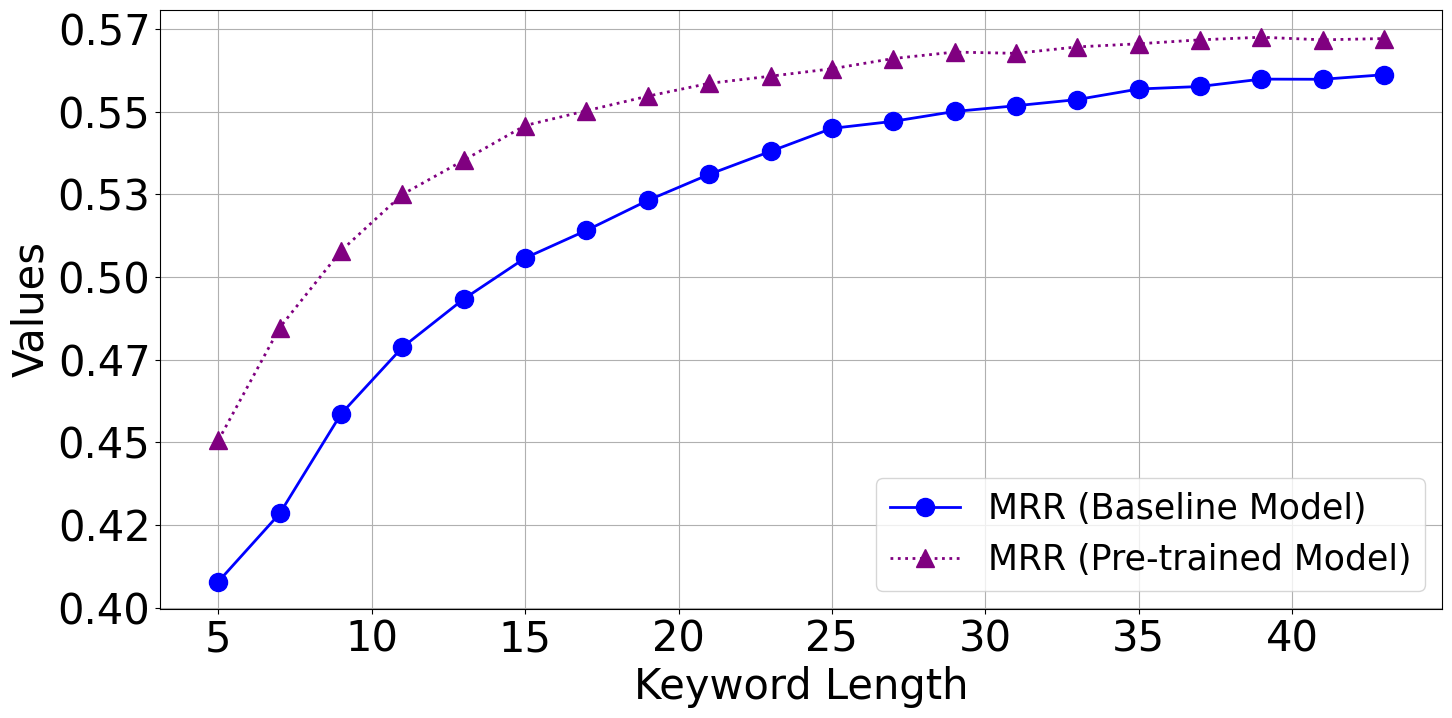}\\
        \small (a) MAP: Baseline vs. Pre-trained & (b) MRR: Baseline vs. Pre-trained\\
    \end{tabular}
    \caption{Choice of Pre-trained, Domain-Specific Embedding Model for Query Reformulation }
    \label{fig:keyword_baseline_pretrained}
\end{figure*}

We also investigate the role of embedding models in our query reformulation step during bug localization, as the quality of reformulated queries directly impacts localization effectiveness. Fig. \ref{fig:codet5_vs_codebert} compares the performance of CodeT5 and CodeBERT in terms of MAP and MRR to determine which model is more effective for generating reformulated queries in our proposed technique. Reformulated queries using CodeBERT achieve a MAP score of 0.306 for a keyword length of 5, which is 27.90\% lower than CodeT5’s score of 0.392. Although increasing the query length reduces the performance gap, CodeBERT does not surpass CodeT5. For instance, at a keyword length of 43, CodeBERT still performs 3.71\% lower than CodeT5. A similar trend is observed for MRR (Fig. \ref{fig:codet5_vs_codebert}(b)), where CodeBERT lags by 27.35\% at a keyword length of 5. Based on these performance differences, we selected CodeT5 for embedding generation in our keyword extraction module, as it leads to more effective query reformulations and thus enhances the overall bug localization performance.

We also wanted to investigate if domain-specific embedding helps in constructing queries. Recognizing the importance of domain-specific pre-training to capture nuanced language features \cite{sun2020sifrank}, we pre-trained the \textit{CodeT5-small} \cite{wang2021codet5} model on a comprehensive dataset containing bug reports (see Section \ref{sec:pretraining_model} for details). This pre-training aimed to equip the model with a deeper understanding of software bugs from Java-based systems. Our findings indicate that the pre-trained model consistently outperforms the baseline model across all metrics, achieving superior performance even with fewer keywords. From Fig. \ref{fig:keyword_baseline_pretrained}, we see that the MAP improves from 0.39 to 0.42, representing a 9.68\% increase over the baseline model using 5 keywords. Similarly, it increases from 0.45 to 0.48 with 15 keywords, indicating a 6.98\% improvement. This trend holds true across various query lengths experimented with, ranging from 5 to 43 for all metrics. Therefore, we choose the CodeT5 model pre-trained with bug reports for our technique - IQLoc.

\begin{tcolorbox}[colback=lightergray,colframe=black,arc=1mm,boxrule=0.5pt]
\textbf{RQ2 Summary:} Reformulated queries improved the performance of IQLoc in bug localization  (e.g., HIT@1 by $\approx$42\%) over baseline queries (i.e., bug report), with improvement maximizing on 15 keywords in a query. Similarly, a language model pre-trained with bug reports enhances the performance of IQLoc (e.g., 9.68\% for MAP) by offering domain-specific embedding, with CodeT5 performing best among the models evaluated.
\end{tcolorbox}


\begin{figure}[!ht]
    \centering
    \begin{tabular}{c} 
        \includegraphics[width=0.7\linewidth]{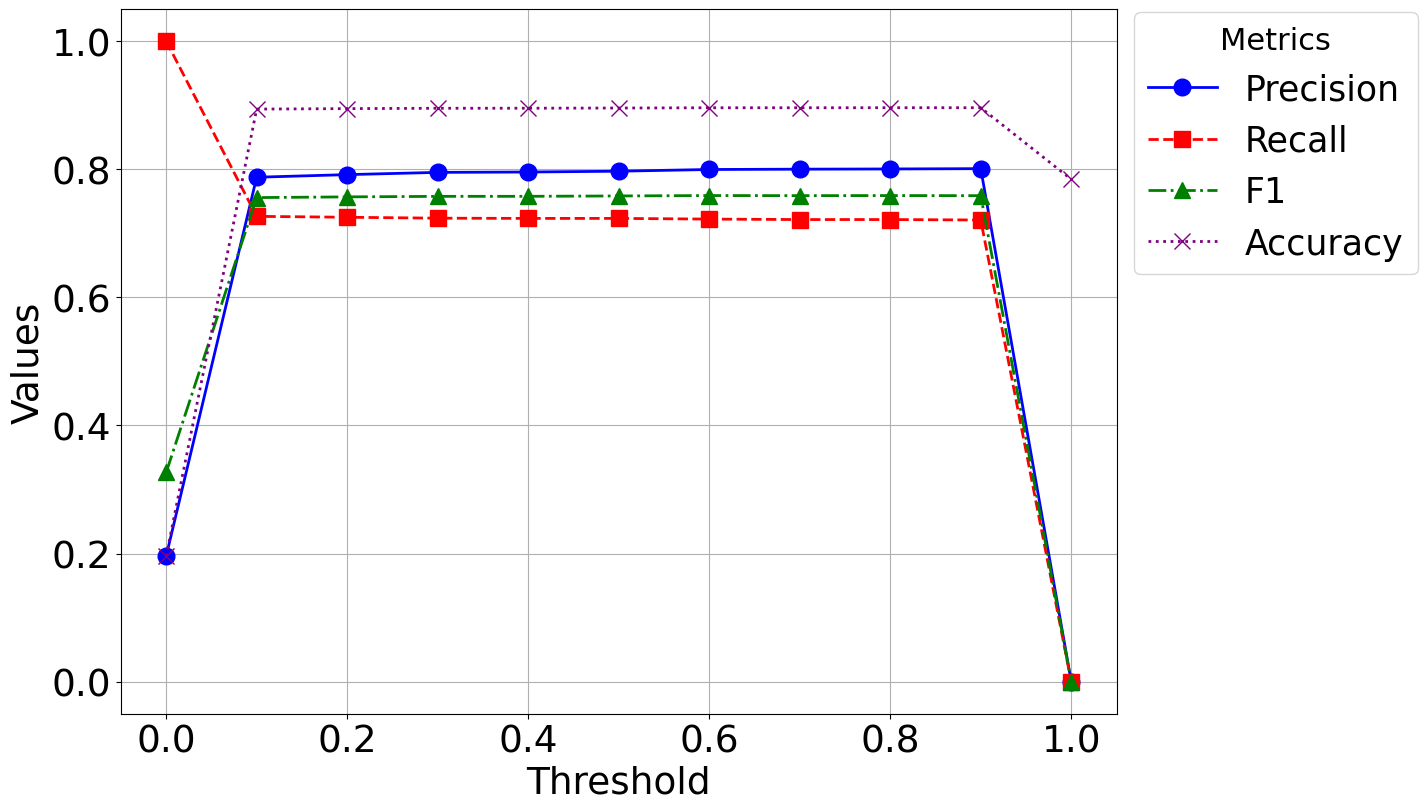}
        \\
        \small (a) CE Performance: Random Split \\
        \\
        \includegraphics[width=0.7\linewidth]{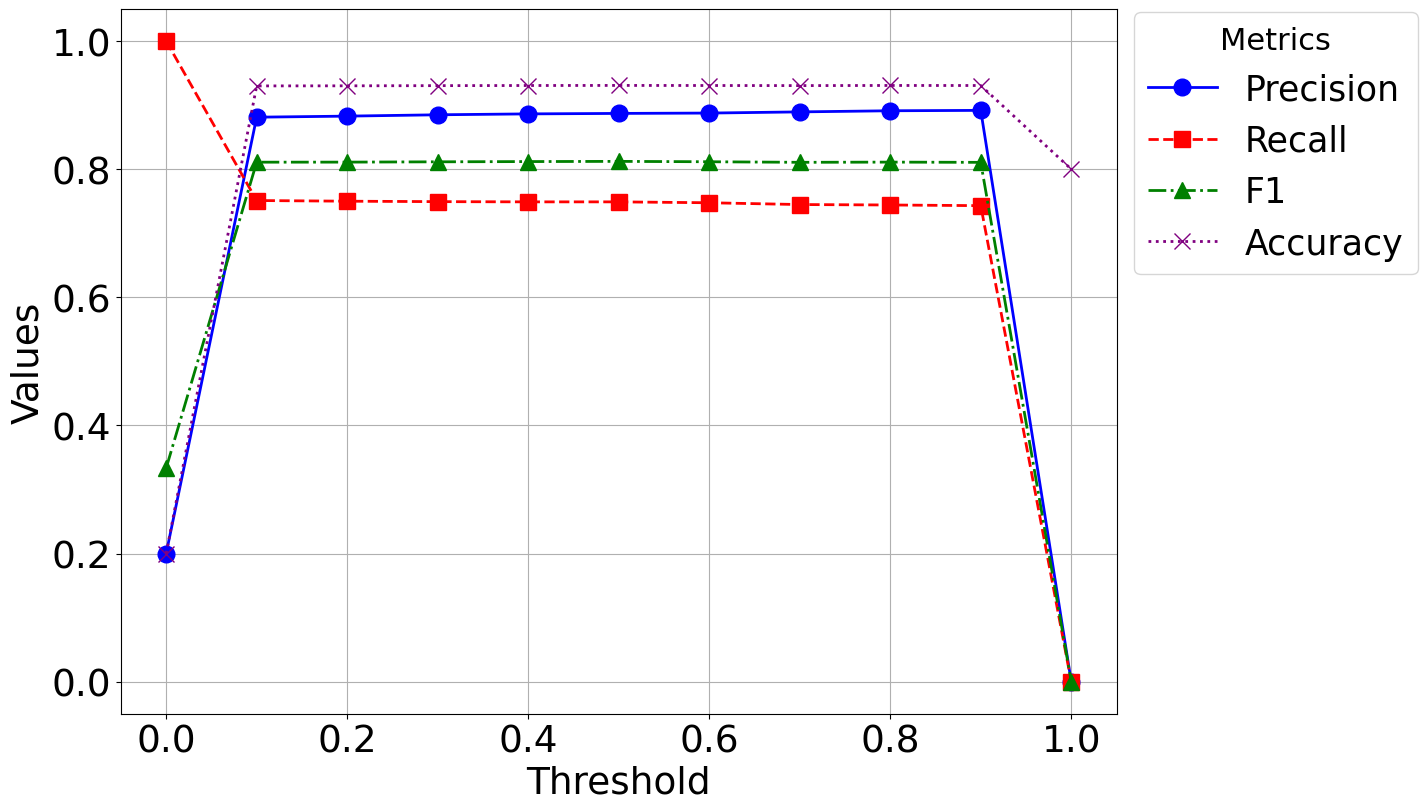}\\
        \small (b) CE Performance: Time-wise Split
    \end{tabular}
    \caption{Cross-Encoder's Performance at Different Relevance Thresholds}
    \label{rq3_crossencoder_performance}
\end{figure}

\subsubsection{Answering RQ$_3$: Performance of the Cross-Encoder Model in Determining Buggy Source Code Based on Program Semantics}

In this section, we investigate the effectiveness of the cross-encoder  model (adopted by IQLoc) in identifying buggy code segments based on their program semantics. Transformer-based cross-encoder models are designed to output probabilistic confidence scores ranging from 0 to 1. We fine-tuned our model to classify code segments as either buggy (i.e., 1) or non-buggy (i.e., 0) against on a given bug report. The model jointly encodes both inputs (i.e., code and bug reports) and learns their contextual relationships during fine-tuning. Fig. \ref{rq3_crossencoder_performance} demonstrates the performance of our adopted cross-encoder model under various configurations, highlighting how changes in threshold affect its ability to leverage learned lexical and semantic features from the bug reports and code segments.

For the random split evaluation set (Fig. \ref{rq3_crossencoder_performance}(a)), at threshold 0, the model achieves the lowest accuracy of 19.5\%, with the lowest precision and F1 scores and the highest recall. This occurs because, at this threshold, the model classifies all instances as positive (a.k.a., buggy), capturing all true positives (TP) but also misclassifying all negative cases as false positives (FP). Since recall is calculated as $TP / (TP + FN)$ and $FN = 0$ in this scenario, recall remains at 1 despite poor precision or accuracy. Once the threshold increases above 0 (e.g., 0.1), performance stabilizes, reaching an accuracy of 91.4\%, with precision, recall, and F1 scores of 81.3\%, 73.8\%, and 75.82\%, respectively. However, at threshold 1, the opposite effect occurs— the model classifies all instances as negative, correctly identifying all negative cases (TN) but misclassifying all positive cases (TP), leading to a recall of 0. Since our evaluation dataset consists of four times more negative cases than positive ones (discussed in Section \ref{section:dataset_constraction}), the accuracy at threshold 1 is approximately 80\%. It is noteworthy that these cross-encoder results for the random split evaluation set are averaged over five independent runs.

In the time-wise split scenario (Fig. \ref{rq3_crossencoder_performance}(b)), our cross-encoder model exhibits similar trends for thresholds 0 and 1. However, once the threshold surpasses 0.1, the model stabilizes and demonstrates slightly improved performance across all metrics. Starting with accuracy, the model shows a consistent upward trend, reaching $\approx$93\%, which is a 1.9\% improvement over the random split dataset scenario. Similarly, precision, recall, and F1 scores remain relatively stable, hovering around 88.7\%, 74.9\%, and 81.2\%, respectively.

The precision of our cross-encoder models at both very high and low thresholds suggests that it makes predictions closer to 1 and 0 for positive (contextually relevant) and negative (contextually not relevant) outcomes, respectively, while maintaining accuracy. Given our focus on the correctness of positive predictions (i.e., precision) while maintaining overall correctness (i.e., accuracy), we set the threshold to 0.5 for our experiments, which achieves the best balance between precision and recall, as reflected in the F1 score.

\begin{tcolorbox}[colback=lightergray,colframe=black,arc=1mm,boxrule=0.5pt]
\textbf{RQ3 Summary:} Our fine-tuned cross-encoder model can effectively identify buggy and non-buggy source code segments leveraging their program semantics and relevance to given a bug report. A threshold of 0.5 ensures an effective separation between the two types of source code by achieving up to 93\% accuracy and 81.1\% precision.
\end{tcolorbox}


\subsubsection{Answering RQ$_4$: Comparison with the Baseline Techniques}

In this section, we compare our proposed technique IQLoc against existing bug localization techniques in terms of various evaluation metrics. In particular, we compare IQLoc with nine baseline techniques -- Baseline Elasticsearch, BLUiR \cite{saha_bleuir}, Blizzard \cite{blizzard}, BRTracer \cite{BRTracer}, BLIA \cite{bl_code_change_BLIA}, LRBL \cite{LRBL_learning2rank}, LLmiRQ \cite{LLmiRQ}, DNNLoc \cite{dnnloc_ir}, and RLocator \cite{RLocator}. 

To replicate the baseline Elasticsearch technique, we indexed all source documents of a subject system's repository and capture the pre-processed bug reports as queries. Then, we execute the queries with the Elasticsearch engine \cite{elasticsearch}, which retrieves the relevant source documents based on the BM25 algorithm \cite{rank-bm25} and Boolean query \cite{boolean_query}. We use the default values for the parameters - k and b - in the BM25 algorithm.

BLUiR \cite{saha_bleuir} employs a structured Information Retrieval approach where it leverages the structural components of both source code documents and bug reports for effective localization, which helps avoid spurious matching. To achieve this, BLUiR collects four code elements (i.e., class names, method names, variable names, and comments) from each source code document and two textual elements from each bug report (i.e., bug title and description). Then, it conducts eight searches to compute suspiciousness score for each code-text pair and then combines scores to calculate the overall suspiciousness score for a source document. For the experiments, we collected BLUiR's replication package from the Bench4BL repository \cite{bench4bl} and adapted it for version-based replication. It uses Indri \cite{Indri}, a TF-IDF \cite{tf-idf}-based search engine, as its backend for experiments, which has become obsolete recently. Thus, we chose to replicate the technique using Apache Lucene \cite{lucene} with a TF-IDF-based scoring algorithm. It should be noted that we adopted the formula suggested by Saha et al.\cite{saha_bleuir} for calculating the TF and IDF metrics, which were then fed to the TF-IDF-based retrieval algorithm of Apache Lucene.

 Blizzard \cite{blizzard} is an IR-based bug localization technique that leverages contextual information from bug reports for query construction. It categorizes bug reports into three types (i.e., NL-Natural Language, PE-Program Element, and ST-Stack Trace), employs three separate graph-based techniques to construct queries from them, and then retrieves buggy source documents from the corpus by executing the queries with the Apache Lucene engine \cite{lucene}. For the replication, we collected and adapted the replication package hosted at GitHub \cite{Blizzard_git_repo} by the original authors.

BRTracer \cite{BRTracer} is an IR-based technique that aims to improve bug localization through document segmentation and stack-trace analysis. Rather than considering each source document as a single unit, it divides the document into equal-sized segments and treats each segment as a corpus document, reducing noise from unrelated parts of large documents. Additionally, it extracts document names from stack traces in bug reports, boosts the ranking of documents appearing in the top 10 stack-trace frames, and assigns a smaller boost score to the remaining related documents. The final ranking combines the segment-based similarity (i.e., rVSM \cite{ir_localization_lda_buglocator}) and the boost scores. We collected the replication package of BRTracer from Bench4BL \cite{bench4bl} and adapted it for our version-based experiments.

BLIA \cite{bl_code_change_BLIA} is another IR-based bug localization technique that combines textual similarity between bug reports and source documents with structured information from code, stack-trace clues, and code change histories. It adopts the revised Vector Space Model (rVSM) for similarity scoring and extends with structured information retrieval \cite{saha_bleuir}. The technique also considers similar past bug reports, boosts documents mentioned in stack traces, and leverages recent commit logs filtered by keywords. These analyses are then integrated with tunable parameters to compute a final suspiciousness score for ranking the source documents. We collected and adapted the BLIA replication package from Bench4BL \cite{bench4bl} for our experiments.

LRBL \cite{LRBL_learning2rank} is a Learning-to-Rank (LtR)-based bug localization technique that employs a data-driven ranking model to localize software bugs. It formulates bug localization as a feature-based learning-to-rank problem, where each bug report–source document pair is represented by six features capturing lexical and API-enriched semantic similarity to the bug report, similarity to previous bug reports associated with the same document (collaborative filtering), the presence of the document’s class name in the bug report, and the document’s historical bug-fixing recency and frequency. Then these pairs are labeled based on historical bug fix data and trained with SVMrank \cite{SVMRank} to rank the suspicious source documents higher in the ranked list. For the experiments, we replicated the technique carefully by following the steps specified by the author.

LLmiRQ \cite{LLmiRQ} is a hybrid technique for bug localization that integrates Large Language Model (LLM)–driven query generation with a supervised Learning-to-Rank (LtR) model. It first classifies each bug report into three categories - program elements (PE), stack trace (ST), and natural language (NL), and uses LLM prompting to reduce search queries for PE and ST reports, and expand queries for NL reports. Using the generated query, it retrieves candidate source documents and trains an SVMrank \cite{SVMRank} model on each (bug report, document) pair with features such as textual similarity, class-name overlap, call-graph relationships, and historical relevance. Then, LLmiRQ employs the trained model to rank potentially buggy documents. We replicated the techniques carefully by following the author's specifications for our experiment.

DNNLoc \cite{dnnloc_ir} is the first technique for bug localization that combines Deep Learning and Information Retrieval. It uses several features -- bug report-source code similarity (rVSM score \cite{ir_localization_lda_buglocator}), class name similarity, collaborative filtering, bug report recency, and bug report frequency to train a deep learning model and then uses the model to localize the bugs. To replicate DNNLoc, we trained appropriate models by extracting features from our training sets, following the authors' suggestions. During bug localization, we used these trained models to predict suspiciousness scores of source code documents and to rank them.

\begin{table}[!ht]
    \centering

    \caption{Comparison between IQLoc and Baseline Techniques in Bug Localization}
    \vspace{-0.5\baselineskip} 
    \caption*{(a) Average Performance Metrics for Random Split}
    \vspace{-0.5\baselineskip} 
        \begin{tabular}{|l|l|l|l|l|l|}
    \hline
        Technique & MAP & MRR & HIT@1 & HIT@5 & HIT@10 \\ \hline\hline
        Baseline Elasticsearch & 0.457 & 0.486 & 0.389 & 0.618 & 0.703 \\ \hline
        BLUiR & 0.470 & 0.504 & 0.393 & \textbf{0.655} & \textbf{0.745} \\ \hline
        Blizzard & 0.480 & 0.510 & 0.410 & 0.648 & 0.735 \\ \hline
        BRTracer & 0.246 & 0.331 & 0.239 & 0.424 & 0.506 \\ \hline
        BLIA & 0.305 & 0.379 & 0.296 & 0.457 & 0.562 \\ \hline
        LRBL & 0.411 & 0.471 & 0.356 & 0.615 & 0.682 \\ \hline
        LLmiRQ & 0.402 & 0.444 & 0.384 & 0.584 & 0.665 \\ \hline
        DNNLoc & 0.311 & 0.322 & 0.249 & 0.416 & 0.508 \\ \hline
        RLocator & 0.478 & 0.511 & 0.419 & 0.652 & 0.726 \\ \hline
        IQLoc & \textbf{0.493} & \textbf{0.520} & \textbf{0.423}	& 0.647	& 0.721 \\ \hline
    \end{tabular}
\vspace{0.2cm}
     
    \caption*{(b) Performance Metrics for Time-wise Split}
    \vspace{-0.5\baselineskip} 
        \begin{tabular}{|l|l|l|l|l|l|}
    \hline
        Technique & MAP & MRR & HIT@1 & HIT@5 & HIT@10 \\ \hline\hline
        Baseline Elasticsearh & 0.471 & 0.496 & 0.396 & 0.618 & 0.714 \\ \hline
        BLUiR & 0.508 & 0.539 & 0.429 & \textbf{0.686} & \textbf{0.786} \\ \hline
        BLIZZARD & 0.494 & 0.525 & 0.419 & 0.670 & 0.753 \\ \hline
        BRTracer & 0.292 & 0.390 & 0.282 & 0.502 & 0.601 \\ \hline
        BLIA & 0.334 & 0.426 & 0.311 & 0.558 & 0.646 \\ \hline
        LRBL & 0.454 & 0.501 & 0.389 & 0.644 & 0.705 \\ \hline
        LLmiRQ & 0.422 & 0.436 & 0.413 & 0.609 & 0.687 \\ \hline
        
        DNNLoc & 0.324 & 0.336 & 0.232 & 0.477 & 0.591 \\ \hline
        RLocator & 0.505 & 0.532 & 0.439 & 0.683 & 0.731 \\ \hline
        IQLoc & \textbf{0.520} & \textbf{0.553} & \textbf{0.466}	& 0.669	& 0.735 \\ \hline
    \end{tabular}
    \label{tab: baseline comparison}
\end{table}

RLocator \cite{RLocator} is a recent IR-based technique for bug localization that incorporates reinforcement learning \cite{reinforcement_learning}, modeling the localization task as a Markov Decision Process (MDP) \cite{markov_decision_reinforcement}. Before applying reinforcement learning (RL), it retrieves source documents from Elasticsearch based on the bug report and filters suspicious documents using an XGBoost \cite{chen2016xgboost} model. It then employs an RL agent based on an actor-critic framework \cite{konda2000actorcritic} that attempts to rank relevant source documents at the top positions, with MAP or MRR as reward signals. We collected the replication package of RLocator from Zenodo \cite{rlocator_replication_zenodo} and adapted it for our study on version-based bug localization while maintaining the original specifications of the authors.

From Table \ref{tab: baseline comparison}(a) (random split dataset), we see that existing techniques such as baseline Elasticsearch achieve a MAP score of 0.457, while BLUiR, Blizzard, and RLocator achieve a comparative MAP score from 0.470 to 0.480, indicating $\approx$2.71\%-4.89\% improvement. Comparatively, BRTracer achieves a much lower MAP score of 0.246. IQLoc, on the other hand, has a MAP score of 0.493, which is 2.71\% to 100.4\% higher than the baseline scores. In other words, our technique can rank the relevant documents higher than irrelevant ones, where the other baseline techniques might struggle. Similar improvements are observed for MRR and HIT@1, with increases of up to 61.49\% and 69.88\%, respectively, indicating promising performance. Although IQLoc exhibits a marginal decrease in HIT@5 and HIT@10 compared to BLUiR, Blizzard, and Rlocator, it demonstrates an improvement of up to 55.52\% and 57.18\% over the baseline Elasticsearch and DNNLoc, BRTracer, BLIA, LRBL, and LLmiRQ measures.

In Table \ref{tab: baseline comparison}(b) (time-wise split dataset), IQLoc's performance also surpasses that of baseline techniques. While Elasticsearch achieves a MAP score of 0.471, BLUiR, Blizzard, LRBL, LLmiRQ, and RLocator achieve comparative MAP scores between 0.422 and 0.508; Similar to random split performance, BRTracers, BLIA, and DNNLoc's scores range from 0.292 to 0.324. In contrast, IQLoc achieves a MAP score of 0.520, representing improvements ranging from 2.36\% to 78.08\% over the baseline measures. Similar improvements are observed for MRR and HIT@1, ranging from 2.59\%-64.58\%, and 6.15\%-100.9\%, respectively. However, like the randomly split case (Table \ref{tab: baseline comparison}(a)), IQLoc experiences a marginal drop in HIT@5 and HIT@10 compared to BLUiR and Blizzard. Nonetheless, it demonstrates improvements of up to 40.25\% and 24.36\% over the baseline measures in these metrics.

\begin{figure}[t]
    \centering
    \includegraphics[width=0.7\linewidth]{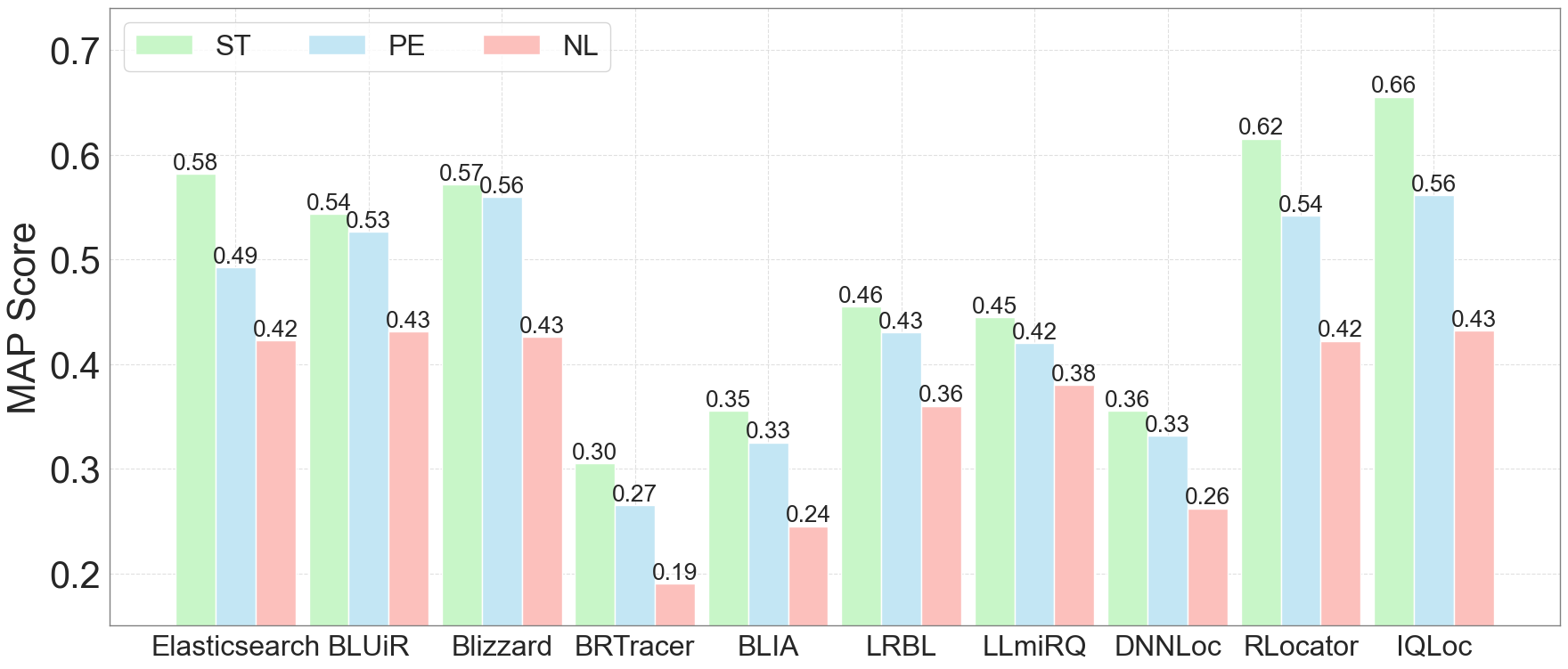}\\[4pt]
    \includegraphics[width=0.7\linewidth]{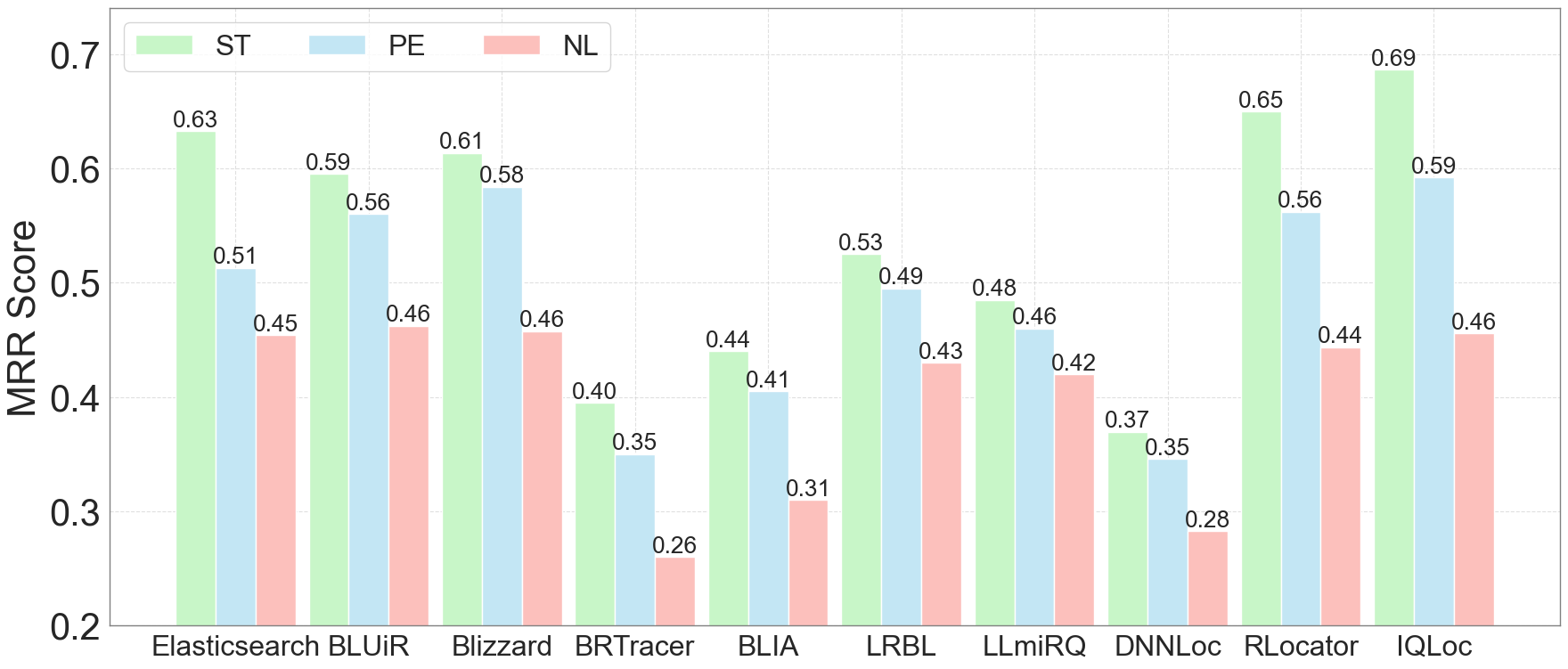}\\[4pt]
    \caption{Comparison of Baseline Techniques in Localizing Different Types of Bugs (Random Split)}
    \label{fig:rq4_random_classwise}
\end{figure}

\begin{figure}[t]
    \centering
    \includegraphics[width=0.7\linewidth]{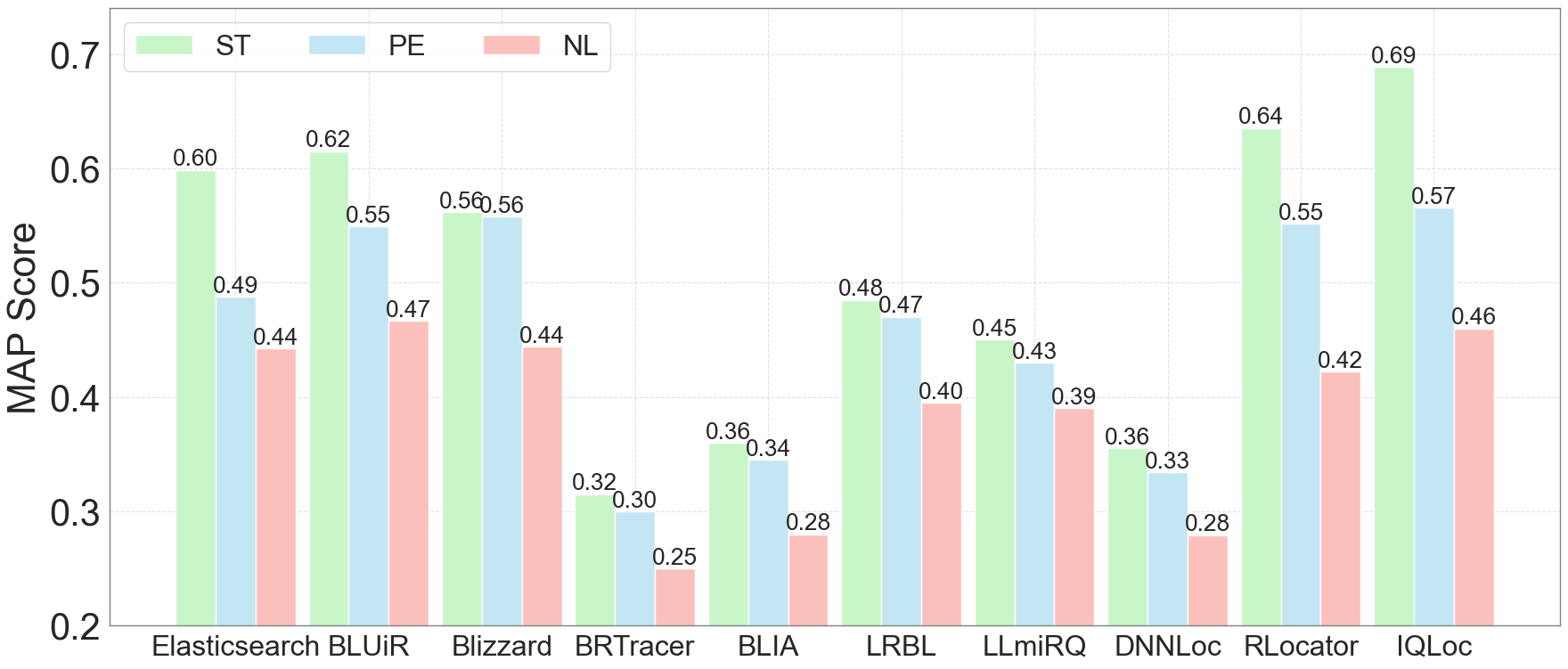}\\[4pt]
    \includegraphics[width=0.7\linewidth]{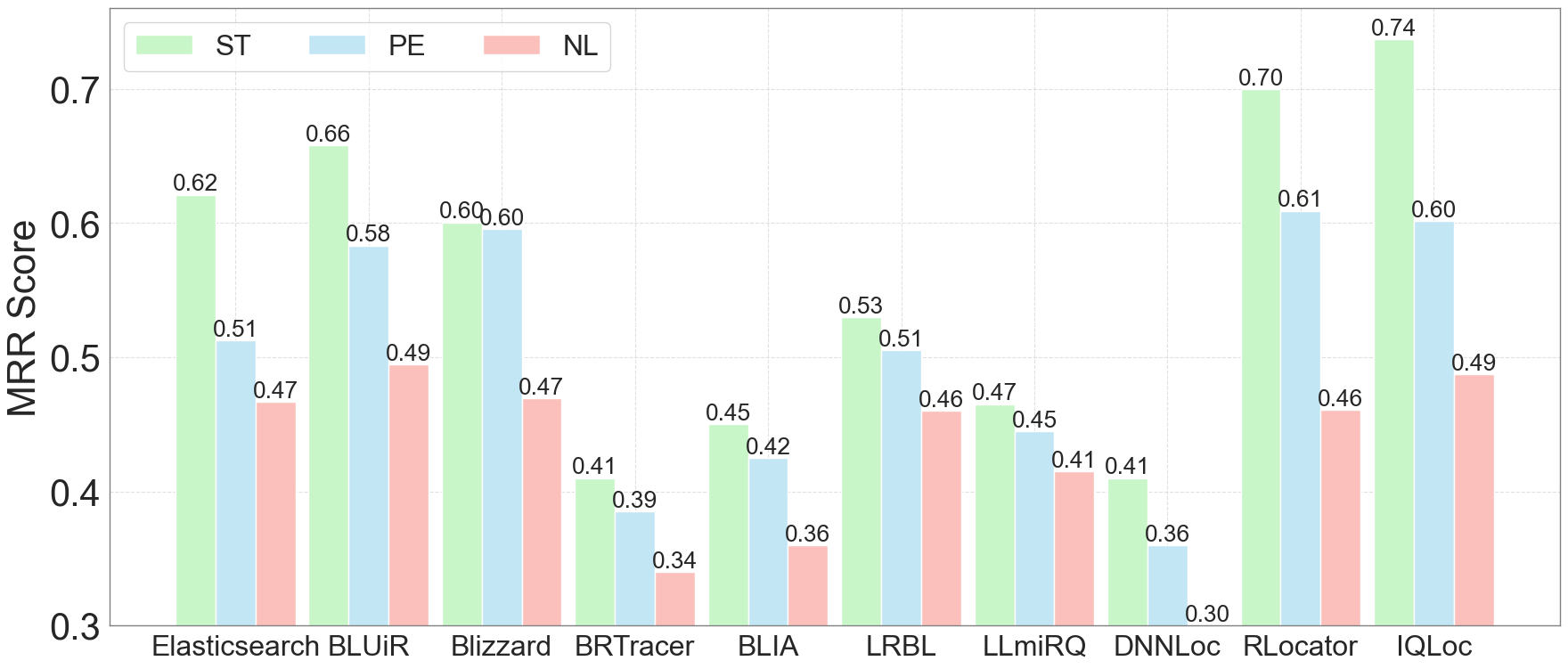}\\[4pt]
    \caption{Comparison of Baseline Techniques in Localizing Different Types of Bugs (Time-wise Split)}
    \label{fig:rq4_time_classwise}
\end{figure}

          
         

\begin{figure}[!t]
    \centering
    \includegraphics[width=0.7\linewidth]{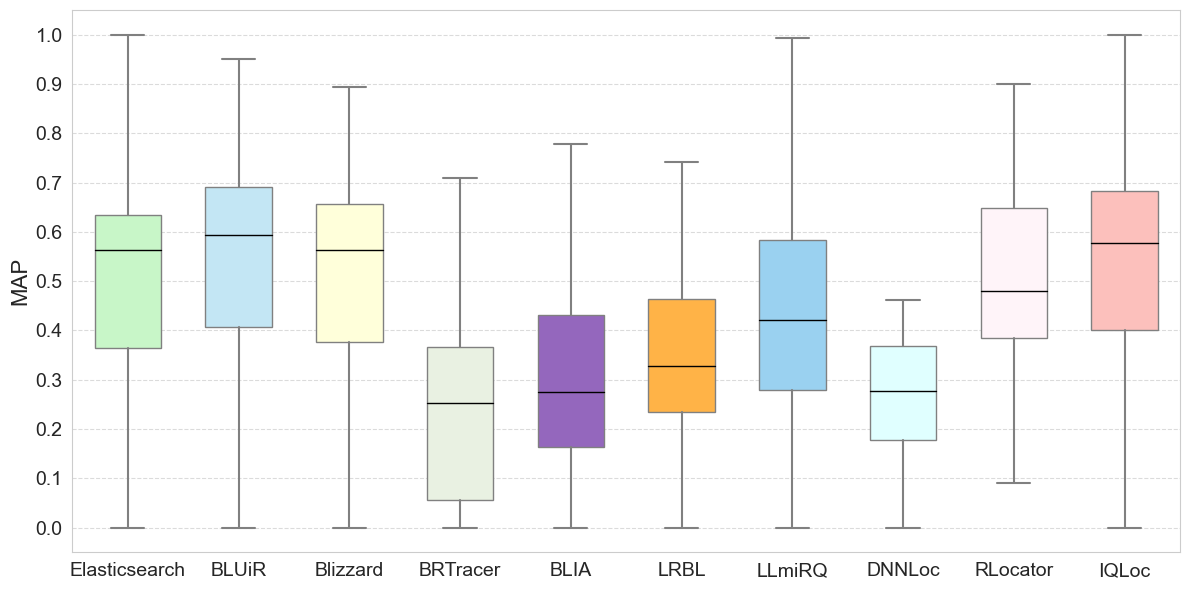}

    \includegraphics[width=0.7\linewidth]{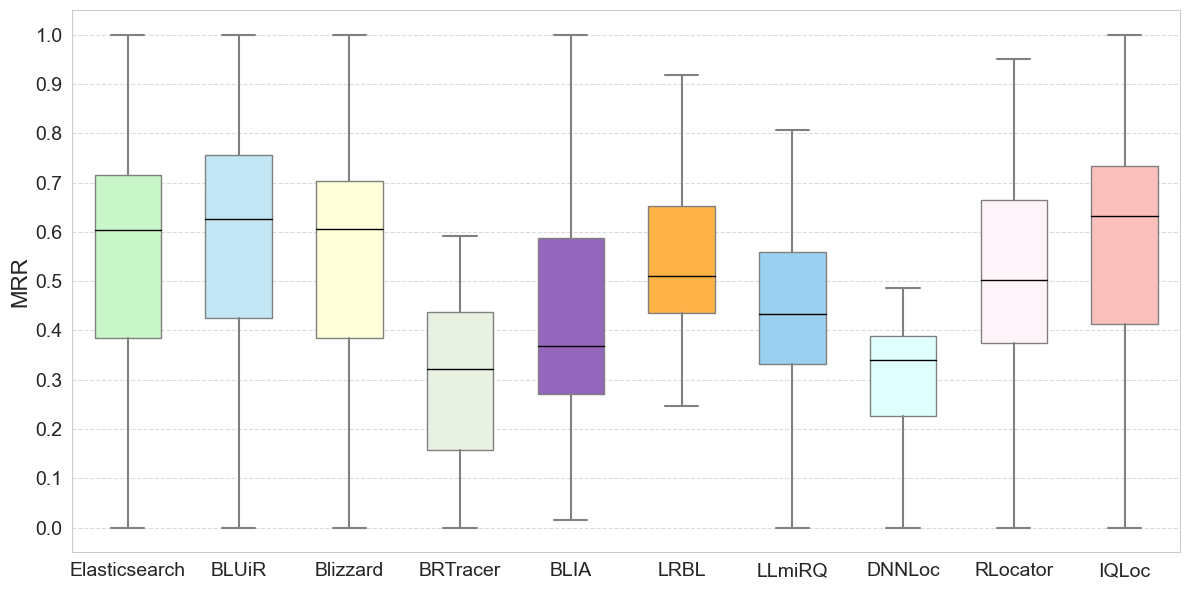}

    \includegraphics[width=0.7\linewidth]{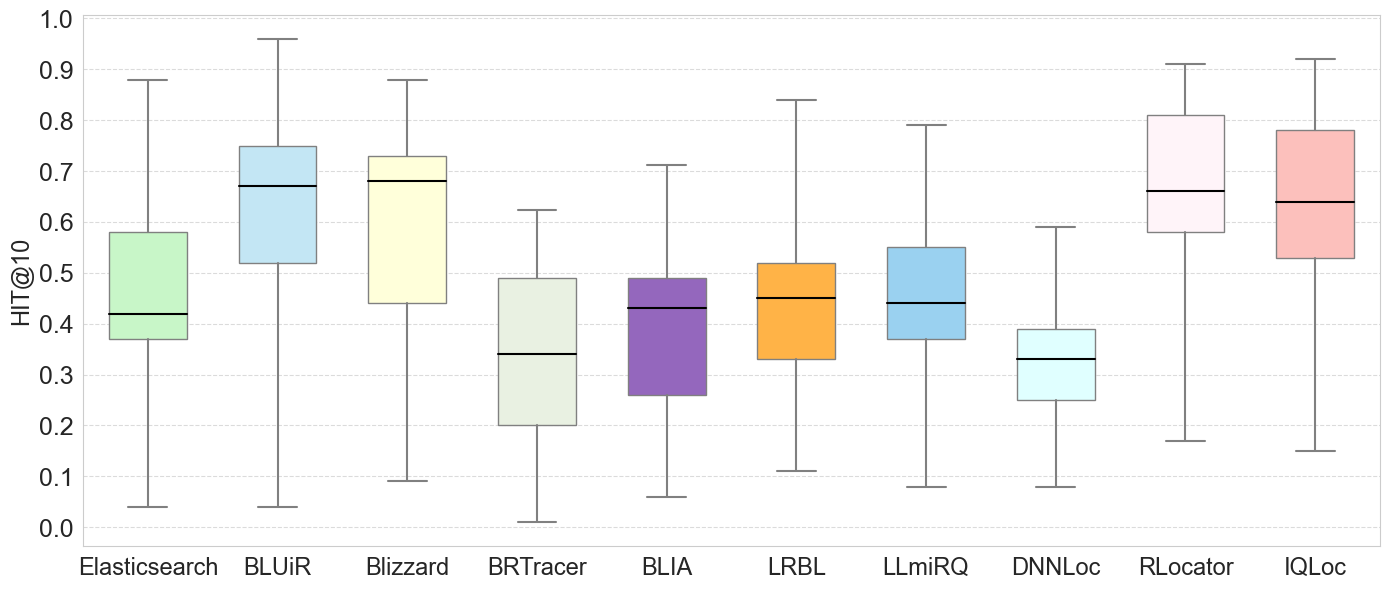}

    \caption{Performance of Different Techniques on Different Subject Systems (Random Split)}
    \label{fig:rq4_random_subjectwise_performance}
\end{figure}

\begin{figure}[!t]
    \centering
    \includegraphics[width=0.7\linewidth]{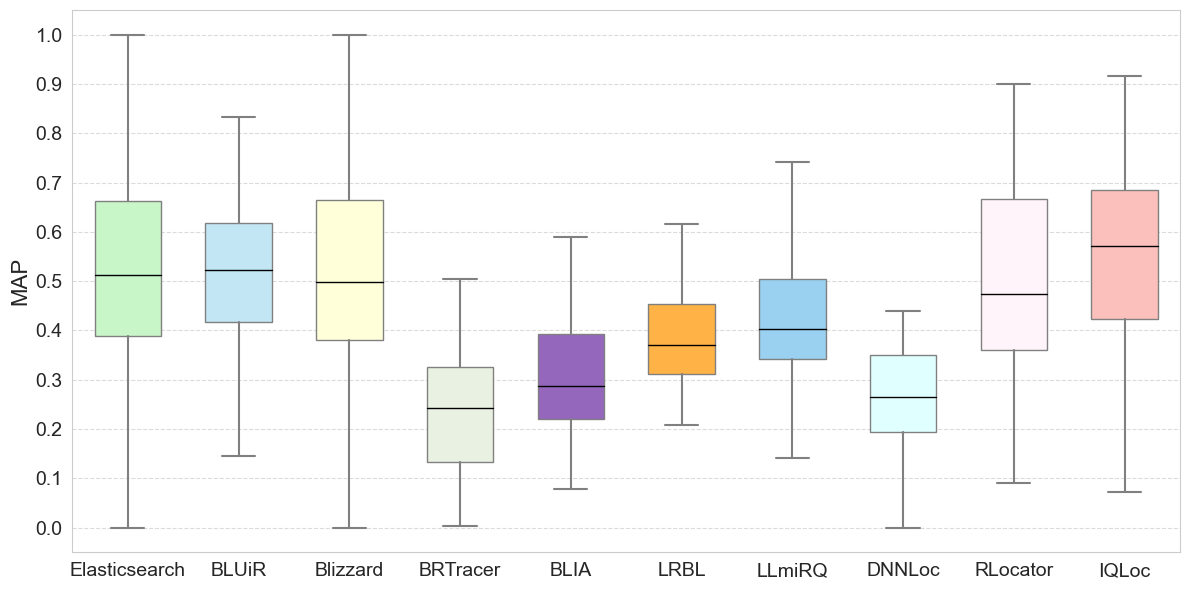}

    \includegraphics[width=0.7\linewidth]{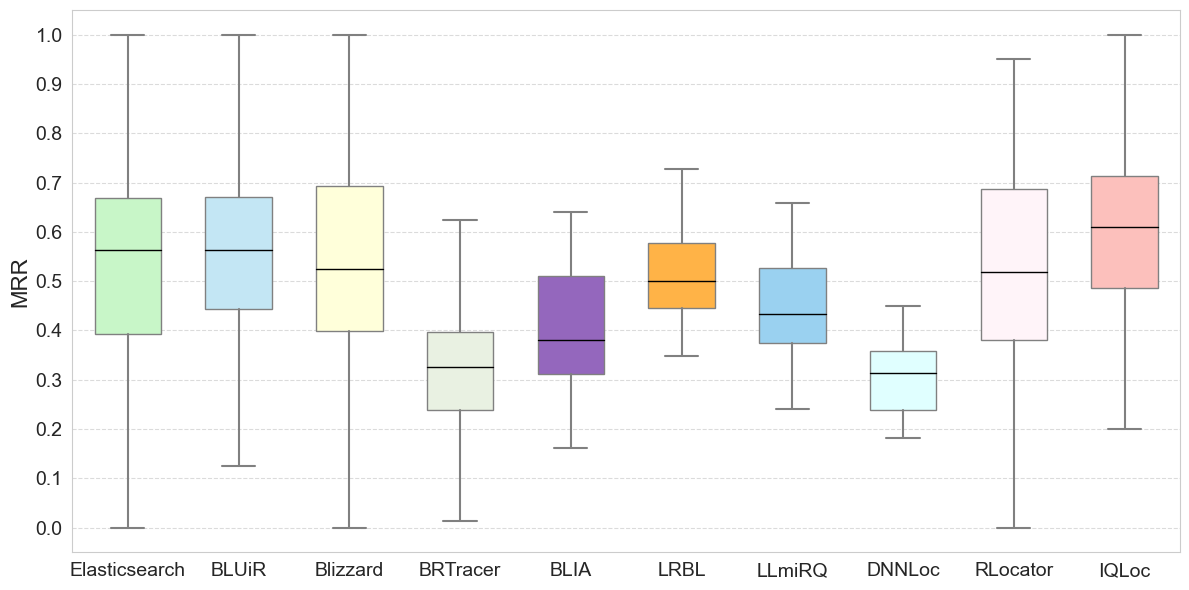}

    \includegraphics[width=0.7\linewidth]{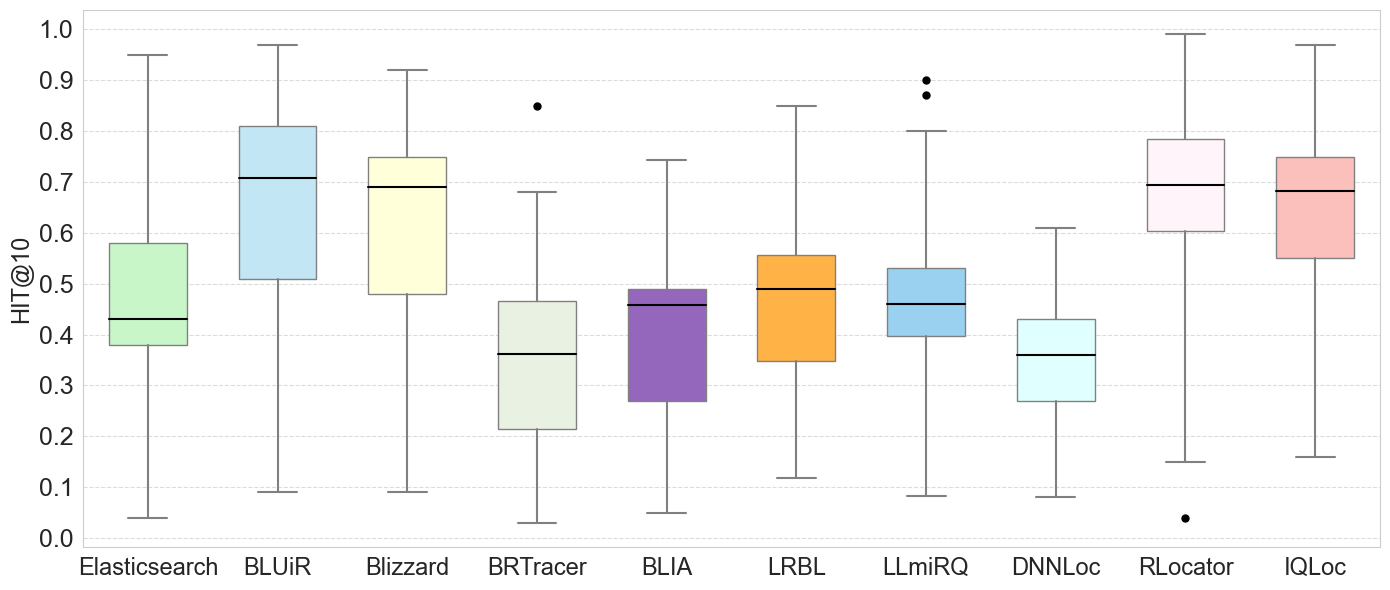}

    \caption{Performance of Different Techniques on Different Subject Systems (Time-wise Split)}
    \label{fig:rq4_timed_subjectwise_performance}
\end{figure}


We also demonstrate the effectiveness of IQLoc in localizing bugs for different types of bug reports and compare it with the baseline techniques. Fig.~\ref{fig:rq4_random_classwise} and Fig.~\ref{fig:rq4_time_classwise} present the performance of various techniques on bug reports containing Stack Traces (ST), Program Elements (PE), and Natural Language (NL)~\cite{blizzard}. For the random-split test set (Fig.~\ref{fig:rq4_random_classwise}), LRBL and LLmiRQ report similar MAP scores ($\sim$0.46) for ST bug reports. Elasticsearch, BLUiR, and BLIZZARD obtain MAP scores in the 0.54-0.58 range, with RLocator achieving a modest improvement at 0.62. IQLoc outperforms all techniques, achieving a MAP 0.66, a 6.42\% improvement over RLocator. For PE bug reports, BLUiR, Blizzard, and RLocator score between 0.53 and 0.56, whereas Elasticsearch lags at 0.49; IQLoc again leads with a MAP score of 0.56. BRTracer, BLIA, and DNNLoc consistently underperform, scoring 0.30 to 0.36 for ST and 0.27 to 0.33 for PE. For NL-only bug reports, all techniques struggle, with IQLoc performing comparably. A similar pattern is observed in MRR, where IQLoc improves by 5.69\%-85.78\% for ST and 1.38\%-71.09\% for PE. For NL-only bug reports, MRR improvements over the baselines reach $\approx$75\%.

A similar pattern is observed in the time-wise split test set (Fig.~\ref{fig:rq4_time_classwise}), where IQLoc achieves an 8.43\%-118.71\% improvement over the baselines for ST and a 2.94\%-88.48\% improvement for PE in terms of MAP. For MRR, IQLoc shows improvements of up to 79.67\% for ST bug reports and 67.02\% for those containing program elements. However, for bug reports containing only natural language, all techniques perform similarly, except DNNLoc, BRTracer, and BLIA, which remain the lowest-performing baseline techniques. 


We further analyze each technique's performance across 42 subject systems in our dataset  and compare their MAP, MRR and HIT@10 performance using box plots (Fig. \ref{fig:rq4_random_subjectwise_performance} and Fig. \ref{fig:rq4_timed_subjectwise_performance}). For the random split test set (Fig. \ref{fig:rq4_random_subjectwise_performance}), IQLoc achieves a higher median MAP than that of all competitors except BLUiR. However, our technique outperforms all baselines in terms of median MRR. For HIT@10, we observe a slight drop in median performance compared to RLocator, but IQLoc still performs better overall than the other techniques.

For the time-wise split test set (Fig. \ref{fig:rq4_timed_subjectwise_performance}), IQLoc outperforms all techniques with a higher median value and a more compact interquartile range (IQR) for MAP, MRR, and HIT@10, indicating lower variability. Although BLUiR exhibits a compact IQR, its median performance remains lower. BRTracer, BLIA, LRBL, LLmiRQ, and DNNLoc, despite showing low variability, consistently perform poorly across all subject systems, reinforcing their lower effectiveness. Overall, IQLoc demonstrates superior performance in localizing bugs across different subject systems with a higher median measure and a reduced variability in performance.

\begin{table}[t]
    \centering
    \caption{Statistical Test: IQLoc vs. Baselines}
    \small
    \begin{tabular}{|l|c|c|c|}
    \hline
        IQLoc \textit{vs} & \textit{p}-value (P@10) & \textit{p}-value (RR@10) & \textit{p}-value (HIT@10) \\ \hline\hline
        Elasticsearch & 0.0098 (Large$^\dagger$) & 0.0139 (Medium$^\dagger$) & 0.0198 (Medium$^\dagger$) \\ \hline
        BLUiR & 0.0176 (Large$^\dagger$) & 0.0296 (Medium$^\dagger$) & 0.0426 (Small$^\dagger$) \\ \hline
        Blizzard & 0.0209 (Medium$^\dagger$) & 0.0328 (Medium$^\dagger$) & 0.0409 (Small$^\dagger$) \\ \hline
        BRTracer & 0.0061 (Very Large$^\dagger$) & 0.0094 (Very Large$^\dagger$) & 0.0111 (Large$^\dagger$) \\ \hline
        BLIA & 0.0082 (Very Large$^\dagger$) & 0.0115 (Large$^\dagger$) & 0.0202 (Large$^\dagger$) \\ \hline
        LRBL & 0.0132 (Large$^\dagger$) & 0.0377 (Medium$^\dagger$) & 0.0294 (Medium$^\dagger$) \\ \hline
        LLmiRQ & 0.0112 (Large$^\dagger$) & 0.0182 (Large$^\dagger$) & 0.0112 (Medium$^\dagger$) \\ \hline
        DNNLoc & 0.0063 (Very Large$^\dagger$) & 0.0091 (Very Large$^\dagger$) & 0.0103 (Very Large$^\dagger$) \\ \hline
        RLocator & 0.0228 (Medium$^\dagger$) & 0.0391 (Small$^\dagger$) & 0.0377 (Small$^\dagger$) \\ \hline
    \end{tabular}

    \vspace{-1pt}
    {\footnotesize
    P@10 = Precision@10, RR@10 = Reciprocal Rank, $^\dagger$ = Effect size (Cliff's $\delta$)
    }
    \vspace{-0.3cm}

    \label{tab:statistical_significance_effect}
\end{table}

IQLoc's performance improvements over the baseline measures demonstrate its effectiveness across both dataset splits. To further validate this efficacy, we conducted \textit{Wilcoxon signed-rank test} \cite{wilcoxon_signed_rank}, a non-parametric statistical test.
We use the SciPy library \cite{scipy} to compute statistical significance of Precision@10 where we capture all buggy documents matching the ground truths within top-10 positions.
For our significance analysis, we computed Precision@10 for each of the 1,501 samples across 42 subject systems in our test set (Section \ref{DS_expansion_train_test}, Table \ref{table:bench4bl_refined}), comparing IQLoc against multiple baseline approaches. 
During the test, we used the \textit{alternative=`less'} parameter (i.e., baseline values are \textit{less} than IQLoc values) to determine if IQLoc consistently achieved higher precision scores within the top-10 results compared to the baseline. In the case of the time-wise split dataset, IQLoc demonstrated \textit{p-values} ranging from 0.0061  to 0.0228 when compared against baseline techniques (Table \ref{tab:statistical_significance_effect}) for Precision@10. These \textit{p-values} are below the significance threshold of 0.05, indicating statistical significance. We observe similar trends for Reciprocal Rank (RR@10) and HIT@10, although the degree of significance varies across metrics. 
Moreover, IQLoc exhibited effect sizes ranging from medium to very large, as measured by Cliff’s $\delta$ \cite{statistical_significance_mann_whitney_cliff}, further explaining the extent of differences in Precision@10. While the effect sizes against RLocator, BLUiR, and Blizzard are smaller in terms of HIT@10, IQLoc still maintains superior performance over other baseline techniques.
Note that we avoid measuring statistical significance for the randomly split dataset, as we perform five independent runs and average the results for our experiment. Given the evidence above, the null hypothesis is rejected, and IQLoc's performance is found to be significantly higher than the baseline measures.

\begin{tcolorbox}[colback=lightergray,colframe=black,arc=1mm,boxrule=0.5pt]
\textbf{RQ4 Summary:} IQLoc outperforms baseline techniques significantly, with MAP improvements of up to 100.4\% for the random split dataset and up to 78.08\% for the time-wise split dataset. Statistical tests confirm the technique’s superiority across different metrics. This improvement stems from its ability to localize various types of bugs across different subject systems. 
\end{tcolorbox}

\section{Related Work}
\label{sec:rel_work_iqloc}
\subsection{IR based Bug Localization}
Bug localization has been a key area of research for decades, driven by the significant impacts and challenges of software bugs \cite{practitioners_bug_localization_study}. There are two primary categories of bug localization: spectra-based and Information Retrieval (IR)-based \cite{ir_bug_localization1}. Spectra-based methods are known for their complexity and limited scalability \cite{spectra1, spectra2}. In our work, we primarily focus on Information Retrieval for bug localization.

Traditional IR-based bug localization methods \cite{saha_bleuir, ir_localization_lda_buglocator, ir_ml_amalgam, wang_amalgam+, blizzard} typically rely on the vector space model (VSM)\cite{vector_space_model}, which analyzes the token overlap between bug reports and source code to identify buggy documents. Several studies have extended VSM by integrating additional contexts, such as bug report history \cite{saha2014effectiveness_bug_rep_history}, code change history \cite{wen2016locus}, or version history \cite{sisman2012incorporating}, into the IR process. For instance, Zhou et al. introduced BugLocator \cite{ir_localization_lda_buglocator} that employs a combined score of a modified VSM score and previous bug fix history to localize bugs. The traditional VSM-based scoring \cite{vector_space_model} method tends to show bias towards longer documents. To address this, they modified the VSM-based scoring and introduced \textit{rVSM} scoring, which enhances the computation of textual relevance between bug reports and code elements. BLUiR \cite{saha_bleuir} captures four types of structural components from the source code (i.e., class names, method names, variable names, and comments) and two components from bug reports (i.e., bug title and bug description). It then forms pairwise combinations of these components to perform eight separate searches using Indri \cite{Indri}, leveraging a modified TF-IDF approach. The final localization score is computed by summing the scores obtained from these individual searches. AmaLgam \cite{ir_ml_amalgam} integrates BLUiR and BugLocator techniques, along with version history inspired by Google by analyzing historical data from version control systems. These components undergo three independent systems to generate rankings before producing the final ranked lists. AmaLgam+ \cite{wang_amalgam+} takes a step further by incorporating stack trace and bug reporter history, alongside AmaLgam's contexts. It ranks files using five components and then returns a final ranked list of source documents for bug localization. Recently, an IR-based technique, PathIdea \cite{chen2021pathidea}, utilizes bug report logs (e.g., log snippets, stack traces) for bug localization. The authors employ a static analysis tool to construct a file-level call graph and reconstruct system execution paths from the logs. To determine the suspiciousness score for each file, they combine the VSM score, a log score that emphasizes files mentioned in the logs, and a path score that highlights files in the execution path. Some IR-based bug localization methods employ more complex mechanisms, such as Latent Dirichlet Allocation (LDA) \cite{ir_localization_lda_buglocator} and Latent Semantic Index (LSI) \cite{ir_bug_localization4}. Nonetheless, simpler methods have shown comparable performance while being more cost-efficient \cite{bench4bl}.

In our work, we employ a BM25-based approach (e.g., Elasticsearch \cite{elasticsearch}) to collect the candidate source documents. Then, our subsequent modules combine the scalability of textual relevance with Transformer-based code understanding grounded in program semantics to retrieve buggy documents at top-ranked positions.

\subsection{Query Reformulation}
Studies by Mills et al. \cite{mills2020relationship} and Rahman et al. \cite{rahman2021forgotten} suggest that poor queries from bug reports may adversely affect the performance of a VSM-based search. Consequently, numerous approaches propose to construct queries to assist developers in their tasks \cite{query_reformulation_tosem, blizzard, haiduc2013automatic, bl_textrank_2}. These studies for query construction fall into two categories: frequency-based and graph-based keyword selection methods \cite{rahman2021forgotten}.

In frequency-based methods, researchers have utilized TF-IDF \cite{tf-idf} and its variants to extract meaningful keywords from both bug reports and source code for use as queries. For instance, Gay et al. \cite{gay2009use} employed relevance feedback (RF) from users to update the query. They implemented Rocchio’s expansion \cite{rocchio_cambridge} method, enhancing query performance by adding terms from relevant documents with increased weights and suppressing or removing terms from irrelevant documents. Haiduc et al. \cite{haiduc2013automatic} proposed a technique that employs a machine learning model trained on 28 query properties to recommend the best reformulation strategy for a given query. These strategies include query reduction, Rocchio expansion, RSV expansion, or Dice expansion, selected based on the query's properties and performance.

Graph-based methods analyze semantic and syntactic dependencies among words to determine their importance. Rahman and Roy \cite{blizzard} applied the PageRank algorithm \cite{pagerank} on constructed graphs to suggest search keywords from various sources. Subsequently, they explored genetic algorithms \cite{lambora2019genetic} for near-optimal search query construction, leveraging the vocabulary in bug reports for effective query building \cite{rahman2021forgotten}. However, their genetic algorithm-based approach is costly and relies solely on the textual relevance between bug reports and source documents.

These techniques typically use statistics and correlations to generate queries from bug reports. However, they often fail to capture the contextual relevance between bug reports and code which limits the effectiveness of query reformulations. In our approach, we address this by leveraging the broader understanding of natural language and program texts from large language models (e.g., CodeT5, CodeBERT) and identifying the most salient terms from both bug reports and code for query construction. 

\subsection{Deep Learning for Bug Localization}
Recent advancements of deep learning in various domains (e.g., natural language processing, and machine translation) have encouraged its application to bug localization. DNNLOC, an early and influential work in the field of localization, is designed to identify potentially buggy documents by learning from multiple features (i.e., rVSM score, class name similarity, collaborative filtering, bug report recency, and bug report frequency) of bug reports and source code. However, its reliance on certain features like bug fixing recency and frequency is available for only 20-40\% of the bugs \cite{ir_localization_lda_buglocator}. So, the unavailability of these features might affect the performance of the model. A recent technique, FBL-BERT \cite{ciborowska_fbl_bert}, incorporates the ColBERT model \cite{khattab2020colbert} for changeset-based bug localization. ColBERT uses a late interaction architecture, independently encoding bug reports and changesets with BERT, followed by efficient vector similarity calculations for relevance estimation. FBL-BERT enhances this by considering different levels of changeset granularity, enabling offline pre-computation of changeset embeddings, and employing a two-stage process of retrieving and reranking for bug localization. Nonetheless, the model may not be as effective in large-scale environments, where big software projects produce thousands of commits a day. Another recent bug localization technique, RLocator \cite{RLocator}, is a first-of-its-kind reinforcement learning (RL) technique for bug localization. The technique attempts to optimize the ranking of a set of documents by employing RL agents based on an actor-critic \cite{konda2000actorcritic} framework with entropy regularization, using a reward signal derived from ranking metrics (e.g., MAP and MRR) for localizing bugs. Other approaches, such as DeepLocator \cite{deeplocator} proposed bug localization using Convolutional Neural Networks (CNN). DreamLoc \cite{dreamloc} addresses the problem of bug localization by proposing a deep and wide architecture. The deep component applies attention mechanisms to capture textual relevance, while the wide component incorporates domain knowledge through a linear model using features such as bug-fixing history and code complexity. However, in addition to technical challenges, these techniques may encounter inefficiencies when dealing with large numbers of documents for bug localization.

Recent work also investigates graph or tree-structured program representations for fault localization \cite{AGFL, GNet4FL}.  AGFL \cite{AGFL} constructs Abstract Syntax Tree (AST) representation from source code and embeds its nodes using Word2Vec \cite{word2vec}. It applies GraphSMOTE \cite{graphSmote} for class balancing and then trains a GCN model \cite{GCN_Graph_Conv_NN} with an attention mechanism to classify faulty nodes. GNet4FL \cite{GNet4FL} preprocesses code by pruning AST nodes, embedding them with Word2Vec and t-SNE, and incorporating failed-test coverage as node features. It then applies GraphSAGE \cite{GraphSAGE} for context-aware aggregation and uses an MLP to generate suspiciousness rankings for program statements.  However, these techniques may underperform because generic Word2Vec embeddings fail to capture source-code semantics. Additionally, limited test coverage can reduce the availability of meaningful execution signals for many AST nodes (e.g., dynamically invoked code). DeepFL \cite{DeepFL} integrates multiple dimensions of fault-diagnosis information (e.g., spectrum-based suspiciousness, mutation-based suspiciousness, code-complexity metrics, and textual similarity features) and learns how these feature patterns correspond to faulty code elements. By grouping related features hierarchically (e.g., merging mutation-based signals before combining with spectrum-based features), the model learns which combinations of signals reliably indicate a bug and produces a ranked list of suspicious methods. A recent approach \cite{combining_error_guessing} combines developers’ error-guessing knowledge with the logical dependencies between methods. These signals are encoded as coverage matrices and fed into a CNN to predict and rank buggy methods.
Another line of work explores multimodal bug-localization models, which integrate different modalities (e.g., textual bug reports, program-execution spectra, source code) within a unified learning framework.
NetML \cite{netml} integrates textual similarity, execution spectra, and suspicious-word features within a model that learns bug and method-specific weights. It further applies network-Lasso regularization on bug-report and method similarity graphs to encourage similar bugs and program elements to share parameters, thereby improving its ability to rank faulty methods. 
DeMoB \cite{demob} adopts a similar multimodal perspective, encoding bug reports and source documents as distinct modalities using an attention-based BiLSTM and a multi-grained dynamic CNN–BiLSTM encoder, respectively. It detects buggy documents by projecting both representations into a shared embedding space and ranking source files according to their learned relevance to the bug report, optimized with a pairwise ranking loss.
However, if training data is limited or execution spectra are unavailable, these techniques may struggle to learn alignments between bug reports and faulty code.

In contrast, our hybrid approach addresses these challenges by focusing on a limited set of documents using Transformer models for scalability and performance. Leveraging the self-attention mechanism of the Transformer model, we process both source code and bug reports simultaneously and analyze their semantics, which helps detect their contextual relevance to buggy code elements. This methodology offers a promising avenue for overcoming the limitations of existing bug-localization techniques by moving beyond surface-level similarity and capturing deeper program-semantic understanding \cite{program_uderstanding}.

\section{Threats to Validity}
\label{sec:threat2val_iqloc}
Threats to \textit{internal validity} relate to experimental errors and biases. Re-implementation of the existing baseline techniques could pose a threat. For BLUiR \cite{saha_bleuir}, Blizzard \cite{blizzard}, BRTracer \cite{BRTracer}, BLIA \cite{bl_code_change_BLIA} and RLocator \cite{RLocator}, we collected the replication packages from Bench4BL \cite{bench4bl}, Rahman et al.’s GitHub repositories \cite{BLIZZARD_github} and from Zenodo \cite{rlocator_replication_zenodo}. However, since the Indri \cite{Indri} library has become obsolete, we replaced that with Apache Lucene \cite{lucene} in the BLUiR replication. For DNNLOC \cite{dnnloc_ir}, LRBL \cite{LRBL_learning2rank}, LLmiRQ \cite{LLmiRQ}, as the replication package is not available, we had to replicate them ourselves. While we acknowledge the potential for implementation errors, we addressed this concern by adhering to the settings and parameters of the original authors and through extensive testing. We also repeat our experiments on two different datasets \cite{blizzard, ir_localization_lda_buglocator} and compare the performance with baselines to mitigate any bias due to random trials. To further mitigate randomness and potential implementation bias, we repeated each random-split experiment five times and report averaged results.

Threats to \textit{external validity} relate to the generalizability of our work. Even though IQLoc is evaluated using only Java code, the underlying models can be adapted to different programming languages through appropriate fine-tuning \cite{domain_adapt}. 

Threats to construct validity relate to the evaluation metrics for our work. We use several metrics such as Mean Average Precision (MAP), Mean Reciprocal Rank (MRR), and HIT@K, which are widely adopted in recommendation systems \cite{blizzard, saha_bleuir, Rack} and Information Retrieval \cite{evaluation_ir, evaluation_metrics}. This confirms no or little threats to construct validity.


Finally, we adapted Bench4BL \cite{bench4bl} dataset for our experiments, which might contain biases \cite{bias_buglocalization} and data quality issues (e.g., misclassified bugs, erroneously labeled buggy files). During the training of the Cross-Encoder model, we accepted method bodies containing buggy lines from Java classes as context. However, the impact of different context sizes was not well tested, which we consider as a scope of future work.

\section{Conclusion and Future Work}
\label{sec:summary_iqloc}
Software bugs account for 50\% of development time and cost billions of dollars annually. Despite significant research over the last few decades, bug localization remains a challenging task for developers. In this study, we introduce IQLoc, a novel hybrid approach that capitalizes on the strengths of both Information Retrieval (IR) and LLM-based program semantics understanding to support bug localization. Our approach enhances IR-based bug localization with reformulated queries, derived from the program understanding of Transformer models. By going beyond surface-level semantic relevance, IQLoc identifies buggy code more accurately. In our evaluation, we refined and extended the benchmark dataset Bench4BL and compared IQLoc against several baselines using three key metrics: Mean Average Precision (MAP), Mean Reciprocal Rank (MRR), and HIT@K. The results show that IQLoc consistently outperforms the baseline techniques, with improvements of up to 100.40\% in MAP, 64.58\% in MRR, and 100.90\% in HIT@K across two types of dataset splits. Further evaluation reveals that IQLoc achieves MAP improvements of 118.70\% for bug reports with stack traces, 111.87\% for those containing code elements, and 127.45\% for those written solely in natural language. All these findings highlight the potential of IQLoc as a robust technique for bug localization that seamlessly integrates the textual relevance of traditional techniques with the program semantic understanding of modern Transformer-based models, setting a new benchmark for performance and reliability in addressing software bugs.

In the future, we plan to investigate how different representations of code structure (e.g., data flow, control flow \cite{data_control_pdg_application}) and various tokenization strategies influence code understanding in bug localization using large language models (LLMs). Moreover, we aim to explore how agentic systems \cite{swe-agent} can help address the multifaceted nature of bug reports in the context of bug localization using Information Retrieval (IR) techniques.






\bibliographystyle{elsarticle-num} 
\bibliography{bibliography} 





\end{document}